\documentclass[10pt,journal,compsoc]{IEEEtran}
\usepackage[utf8]{inputenc}
\usepackage{multirow}
\usepackage{array}

\usepackage{bbding}
\usepackage{graphicx}  
\usepackage{tablefootnote}
\usepackage{flushend}
\usepackage{subfigure}
\usepackage{amsmath,amsfonts}
\usepackage[colorlinks,linkcolor=black]{hyperref}

\ifCLASSOPTIONcompsoc
  \usepackage[nocompress]{cite}
\else
  \usepackage{cite}
\fi

\ifCLASSINFOpdf
\else
\fi

\begin{document}
%
\title{More Like Real World Game Challenge for Partially Observable Multi-Agent Cooperation}
%
%
%
%

\author{Meng Yao,
        Xueou Feng,
        Qiyue Yin*
\IEEEcompsocitemizethanks{\IEEEcompsocthanksitem Meng Yao, Xueou Feng and Qiyue Yin are
with Institute of Automation, Chinese Academy of Sciences, Beijing,
China, 100190; Qiyue Yin is also with University of Chinese Academy of Sciences,
Beijing, China, 100049.\protect\\
E-mail: meng.yao@ia.ac.cn, xueou.feng@outlook.com, qyyin@nlpr.ia.ac.cn
\IEEEcompsocthanksitem * Corresponding author: Qiyue Yin
}
}

%
%

\markboth{Journal of \LaTeX\ Class Files,~Vol.~14, No.~8, August~2015}%
{Shell \MakeLowercase{\textit{et al.}}: Bare Demo of IEEEtran.cls for Computer Society Journals}
%



\IEEEtitleabstractindextext{%
\begin{abstract}
Some standardized environments have been designed for partially observable multi-agent cooperation, but we find most current environments are synchronous, whereas real-world agents often have their own action spaces leading to asynchrony. Furthermore, fixed agents number limits the scalability of action space, whereas in reality  agents number can change resulting in a flexible action space. In addition, current environments are balanced, which is not always the case in the real world where there may be an ability gap between different parties leading to asymmetry. Finally, current environments tend to have less stochasticity with simple state transitions, whereas real-world environments can be highly stochastic and result in extremely risky. To address this gap, we propose WarGame Challenge (WGC) inspired by the Wargame. WGC is a lightweight, flexible, and easy-to-use environment with a clear framework that can be easily configured by users. Along with the benchmark, we provide MARL baseline algorithms such as QMIX and a toolkit to help algorithms complete performance tests on WGC. Finally, we present baseline experiment results, which demonstrate the challenges of WGC. We think WGC enrichs the partially observable multi-agent cooperation domain and introduces more challenges that better reflect the real-world characteristics. Code is release in http://turingai.ia.ac.cn/data\_center/show/10.

\end{abstract}

}

\maketitle

\IEEEdisplaynontitleabstractindextext

%
\IEEEpeerreviewmaketitle

\IEEEraisesectionheading{\section{Introduction}\label{sec:introduction}}
\IEEEPARstart{M}{any} real-world problems can be modeled as the cooperation of multiple agents, such as self-driving cars \cite{candela2022transferring} \cite{li2022v2x} \cite{zhou2022multi} \cite{toghi2022social}  and multi-robot control  \cite{bettini2023heterogeneous} \cite{johnson2022multi} \cite{liang2022multi} \cite{bettini2022vmas} \cite{qiu2023multi}.  Multi-agent reinforcement learning (MARL) has gained increasing attention for its applications in various domains, and promising performances have been achieved \cite{berner2019dota}  \cite{jaderberg2019human} \cite{vinyals2019alphastar}. Game environments play a critical role in promoting the development of MARL, enabling new MARL algorithms to be tested quickly, safely, and reproducibly. Recently, researchers have focused on partially observable and cooperative settings of MARL to meet real-world requirements.
As one of the most successful environments, the StarCraft Multi-Agent Challenge (SMAC) \cite{smac} focuses on providing standard benchmark for partially observable, cooperative multi-agent problems. Based on this environment, many influential works have been verified. However, in our survey of this field, we find that most current environments are:
\begin{enumerate}[]
\item synchronous, where agents execute actions at the same pace. In reality, heterogeneous agents usually have their own action spaces, and there is no guarantee that actions from different agents will be executed at the same cycle, leading to asynchronous multi-agent cooperation.
\item fixed in the number of agents (leaving out agent death), where action space is limited and not very scalable. In reality, the changeable number of agents requires a flexible action space. 
\item with less stochasticity, where state transitions are simple and safe. However, the real world is often more stochastic and extremely high risk, this means that even when the state is fully observable, the state transition is uncertain.
\item balanced, where both sides of the confrontation have relatively similar abilities. However, in the real world, there is often an ability gap between the two sides of confrontation, leading to an asymmetry. 
\end{enumerate}

\begin{figure}
   \begin{center}
   \includegraphics[width=0.45\textwidth,height=0.6\textwidth]{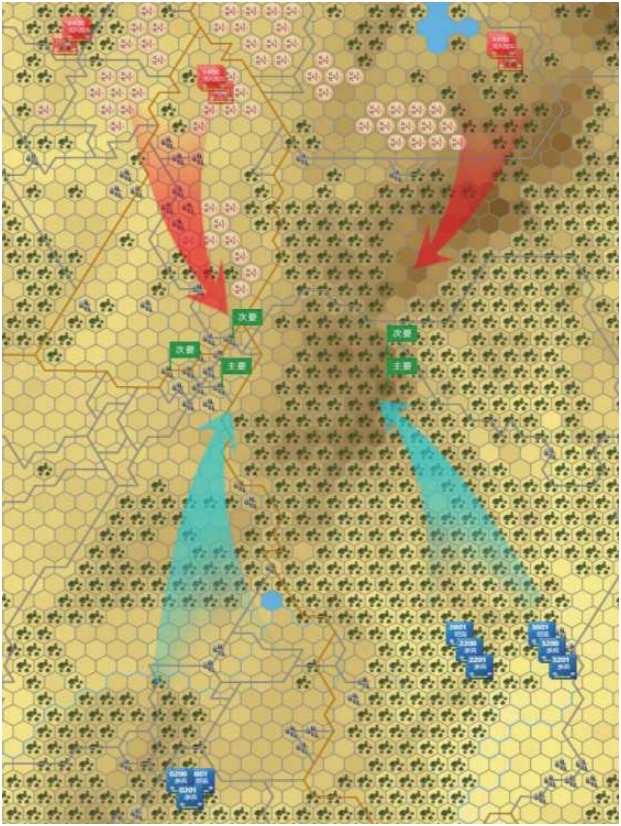}
   \end{center}
   \caption{A game instance of wargame.}
   \label{bq}
\end{figure}

On the other hand, we have also noticed that these issues may pose challenges to existing multi-agent reinforcement learning algorithms. Asynchronous problems require more accurate credit assignment, while high stochasticity can increase the variance of the action value function. Additionally, scalable action space and asymmetry issues also challenge the existing algorithms. These issues have not yet received sufficient attention in MARL benchmark environments we investigated.
Fortunately, we have found that existing games can alleviate these issues. Wargame, a complex confrontational game between two armies abstracted from real-world environment, possesses the characteristics of asynchrony, scalability, strong stochasticity, high risk, and asymmetry. A game instance of Wargame can be seen in the screenshot displayed in Figure \ref{bq}.

Inspired by the Wargame, we propose the WarGame Challenge (WGC), which provides a new standard environment for partially observable multi-agent cooperation to sovle the gap we mention above. Using a complete Wargame as a benchmark for testing MARL algorithms is impractical. It requires a significant amount of effort to acquire skills, such as exploring key points in a large map, and demands extensive computing resources. Therefore, we have abstracted WGC from Wargame by retaining its main features while omitting some specific domain knowledge. WGC comprises five sub-environments and introduces four tasks for MARL that significantly enhance partially observable multi-agent cooperation and are more in line with the characteristics of real-world game scenarios. These tasks include: 1) partially observable asynchronous multi-agent cooperation task, 2) changeable agents multi-agent cooperation task, 3) asymmetric multi-agent cooperation task, and 4) strongly stochastic and high-risk multi-agent cooperation task. 

WGC is a lightweight, flexible, and easy-to-use environment with a clear framework that can be easily customized to meet different experimental requirements. We provide three game scenarios of varying difficulties for each sub-environment, with built-in rule-based AI as opponents. Furthermore, we offer MARL baseline algorithms, such as QMIX, self-play mode, and human-AI mode, for engineering needs to help the algorithm complete the performance tests on WGC. With its configurable nature, WGC has the potential to provide a diverse range of testing environments. Finally, we present the experimental results of knowledge AIs and QMIX on WGC to validate the challenges of WGC.

The rest of the paper is organized as follows. In Section
2, we provide a brief description of typical multi-agent environments. In Section 3, we introduce the WGC environment in detail. In Section 4, we discuss the experiments conducted on WGC. Finally, we conclude the paper in Section 5.

\section{Related Work}
In our investigation of multi-agent environments across different domains, we have found that multi-agent reinforcement learning (MARL) research faces a diverse range of problems.\cite{gronauer2022multi} systematically enumerated challenges that exclusively arise in the multi-agent domain. Currently, it may be challenging to provide a comprehensive environmental benchmark for MARL, similar to gym in single reinforcement learning, in which multiple scenarios can be unified with a general framework.

We provide a list of classic and commonly used multi-agent environments in recent years.
multiple grid world-like environments \cite{lowe} present a set of simple grid-world like environments for MARL with the implementation of MADDPG, which includes cooperative and competitive tasks with a focus on shared communication and low-level continuous control.
\cite{resnick2018pommerman} proposes a multi-agent environment based on the game Pommerman, which consists of a series of cooperative and adversarial tasks.
The StarCraft Multi-Agent Challenge (SMAC) \cite{smac} is a representative and challenging multi-agent game that has been used as a test-bed for various MARL algorithms. Similarly, the recently proposed Google Research Football \cite{kurach2019google} environment has also been used as a test-bed for MARL algorithms.
We have also taken notice of the Fever Basketball \cite{jia2020fever} benchmark, which is an asynchronous cooperation sports game environment that provides perfect information for every agent.
\cite{baker2019emergent} proposed a physics-based environment in which
agents are tasked with competing in a two-team hide-and-seek game.
Many algorithms use those classic environments for testing performance \cite{gronauer2022multi}. SMAC also shows performance test of classic algorithms such as QMIX \cite{rashid2018qmix}, QTRAN \cite{son2019qtran} and COMA \cite{foerster2018counterfactual}. Some well-known and large-scale MARL game environments, such as Dota \cite{berner2019dota}, have played a significant role in promoting the development of distributed MARL algorithms.

Otherwise, considering the different issues in MARL, \cite{yang2018mean} proposed a framework for creating grid worlds covering hundreds to millions of agents with relatively simple game rules. In addition, \cite{leibo} developed a mixed-cooperative Markov environment to test social dilemmas, while \cite{peng2021facmac} created the Multi-Agent MuJoCo (MAMuJoCo) environment for more complex continuous domains to stimulate more progress in continuous MARL. Environments for traffic topics \cite{zhang2019cityflow} and multiplayer competition games such as Doudizhu \cite{zha2021douzero} have also received widespread attention.

Most of the environments mentioned above have not focused much on the four tasks we mentioned. However, we believe that investigating these four problems has significance for understanding real-world game and discovering new algorithms, and this is also why we proposed the WGC environment.

\begin{figure*}[h!]
\centering
\subfigure[No special terrain ]{  
 \includegraphics[width=2 in]{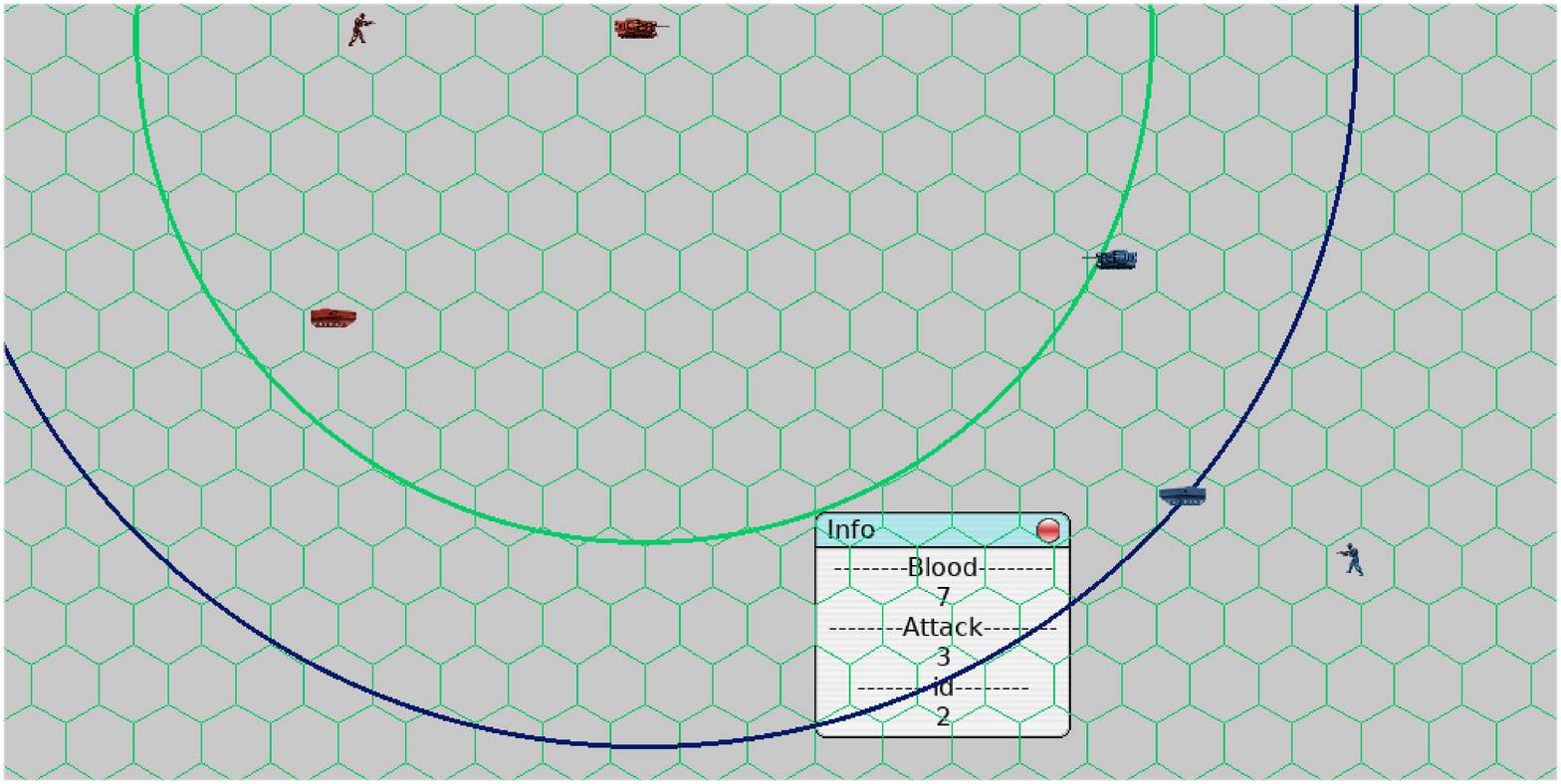}
}
\subfigure[3V3]{  
\includegraphics[width=2 in]{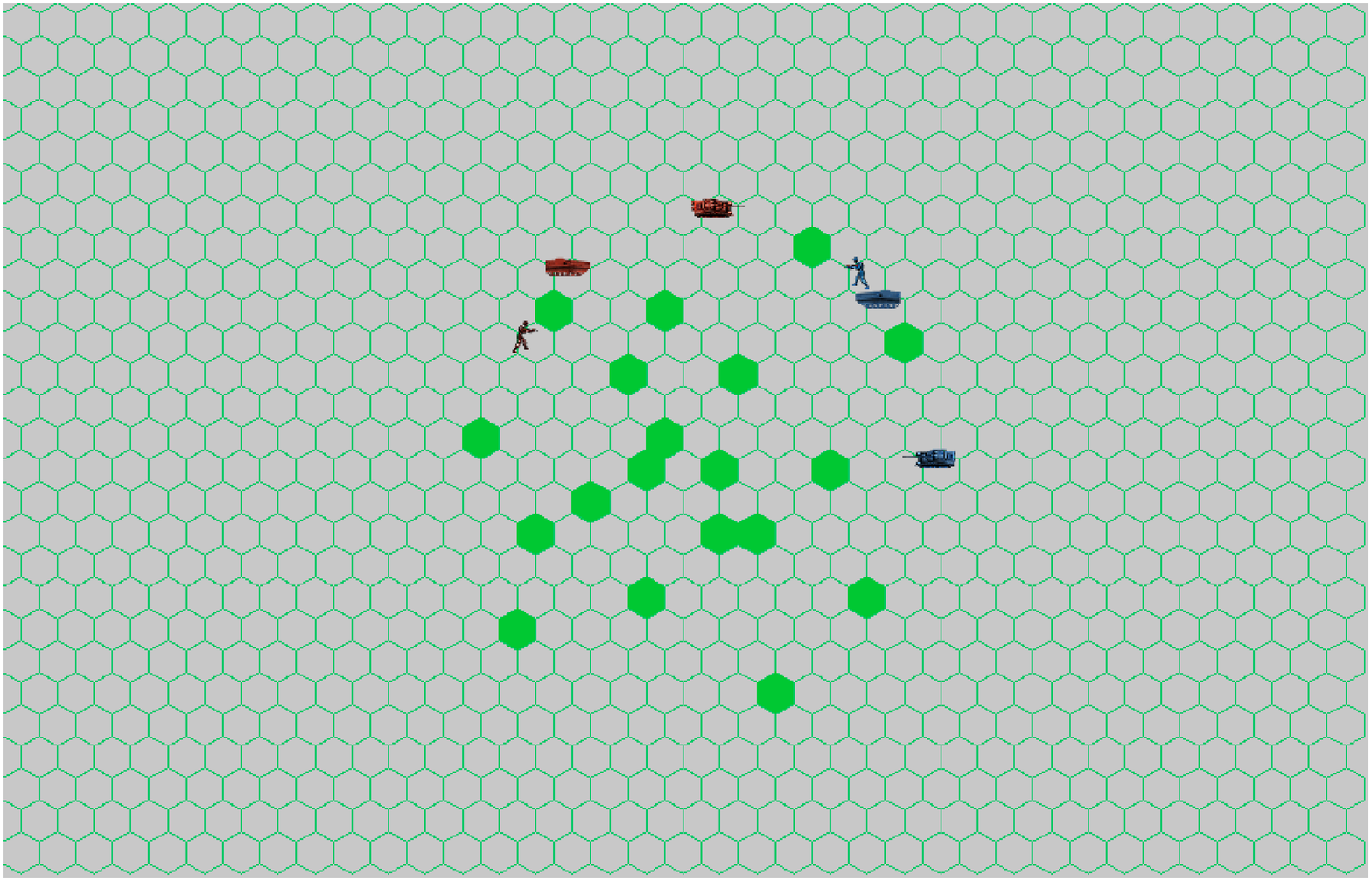}
}
\subfigure[5V3]{  
\includegraphics[width=2 in]{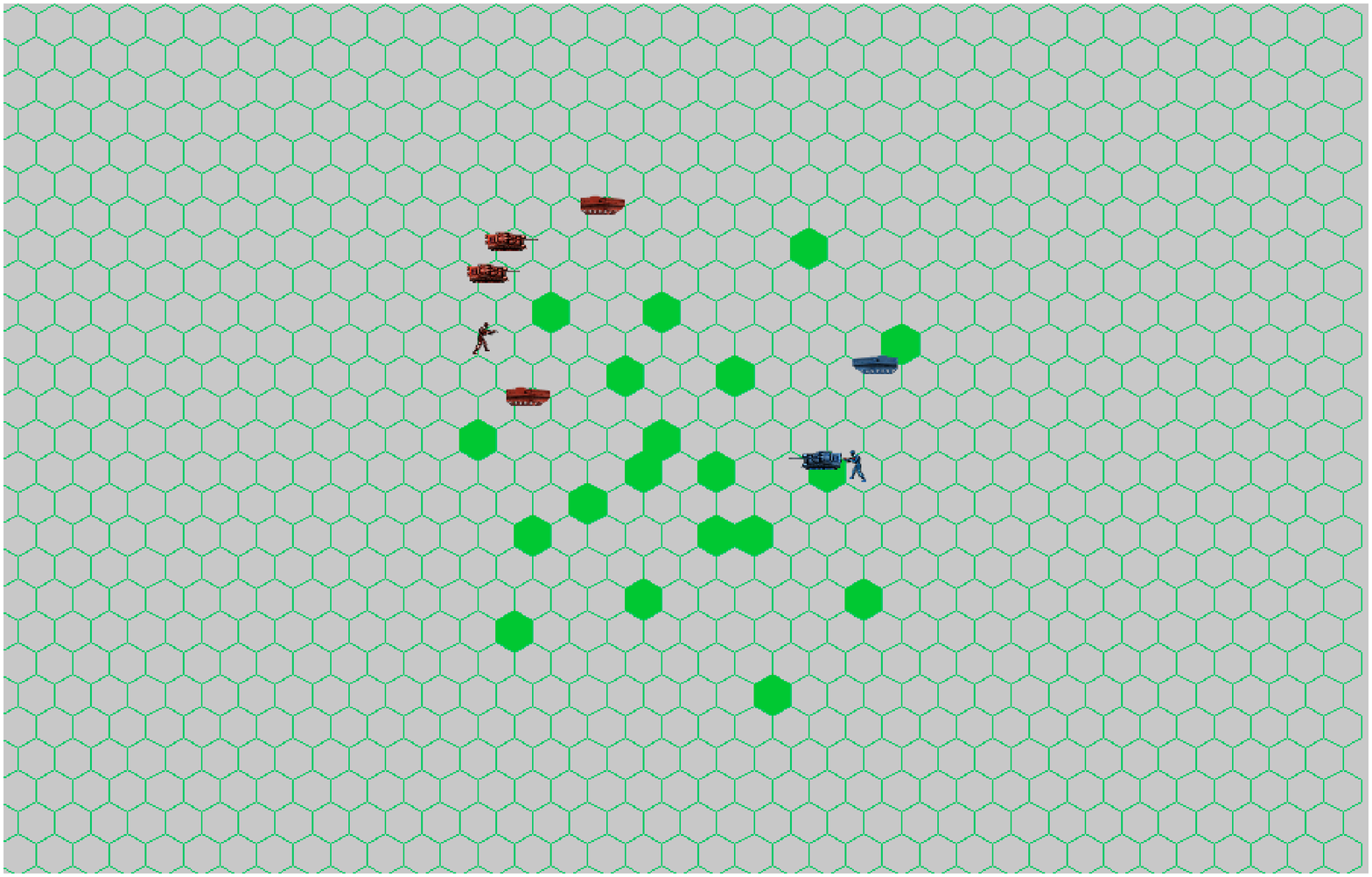}
}
\caption{Scenario example of WGC. }    
\label{biqi}       
\end{figure*}

\section{WarGame Challenge}
War Game, also known as wargame, is a classic strategy game that has garnered widespread attention due to significant military impact and its similarity to real-world environments. In recent years, researchers have dedicated their efforts to solving the wargame challenge by developing military intelligence\cite{moy2019application}. In this paper, we have simplified the wargame challenge to create a universal environment called the WarGame Challenge (WGC), an illustration of our environment is presented in Figure \ref{biqi}. The WGC provides targeted research scenarios by creating five sub-environments: 
\begin{enumerate}[]
\item The Standard sub-environment is designed for the partially observable multi-agent cooperation task, also known as the standard Dec-POMDPs problem. 
\item Partially observable asynchronous multi-agent cooperation sub-environment (POAC) is designed for partially observable asynchronous multi-agent cooperation task. 
\item Partially observable changeable agents multi-agent cooperation sub-environment (CMAC) is designed for partially observable changeable agents multi-agent cooperation task. 
\item Partially observable asymmetric multi-agent cooperation sub-environment (AMAC) is designed for partially observable asymmetric multi-agent cooperation task.  
\item Strongly stochastic and high-risk multi-agent cooperation sub-environment (SRMAC) is designed for partially observable strongly stochastic and high-risk multi-agent cooperation task. 
\end{enumerate} 

All of the sub-environments in the WarGame Challenge share a set of fundamental elements. However, each sub-environment also has its own unique supplementary elements. In this section, we will first introduce the shared elements that are common to all sub-environments. Next, we will discuss the distinctive characteristics of each sub-environment, and finally, we will introduce the overall framework of the WarGame Challenge.

\subsection{Shared Elements}
In wargame, operators from the red and blue teams engage in combat on a specific map while adhering to pre-determined rules. Similarly, the WarGame Challenge (WGC) revolves around three main elements for sub-environments: operators, maps, and rules. The built-in AI bot and scenarios, based on the three main elements, are also important elements of the sub-environments. Otherwise, WGC provides shared elements for multi-agent reinforcement learning. In this section, we will introduce all elements of WGC.

\subsubsection {Main Elements For Sub-Environment}
Operator, map and rule are three main elements of WGC shared by all Sub-Environments. On this basis, scenarios and built-in AI bot emerge, and due to the sharing of elements, it is relatively easy to obtain sub-environments with different functions through modification.

\textbf{Operator}.
There are three types of operator in the WGC: chariot, tank, and infantry. A general summarization of operator attributes for all types is provided in Table \ref{att}. When the attribute values of the operators change, the environment also undergoes significant changes. We have created a template for illustration, which is presented in Table \ref{tab:tank}. The template demonstrates that the tank has the highest health, no shoot preparation time, a higher probability of causing damage in an attack, and ability to move quickly. As a result, the tank is well-suited for charging. The chariot, on the other hand, can inflict the most damage to a tank, but its shoot preparation time makes it necessary to be protected. Lastly, the infantry moves slowly but loses less health when attacked. Its observed distance is so short that it can not be detected by enemies easily, allowing it to hide while searching for enemies and cooperate with the chariot. Users can also adjust operator attribute parameters to create more diverse gameplay.

\begin{figure*}
   \begin{center}
   \includegraphics[width=0.7\textwidth]{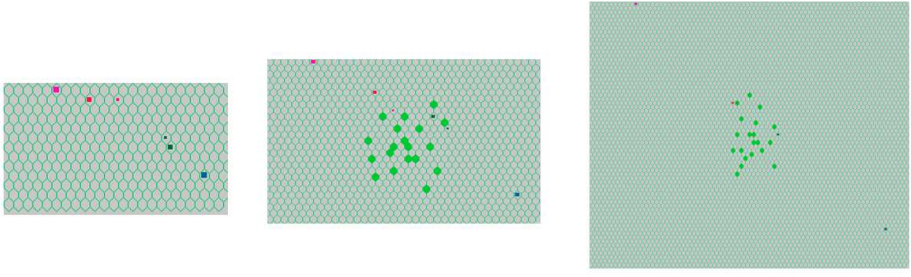}
   \end{center}
   \caption{Map data is shared across all sub-environments. The left  map is the smallest map with no terrain, the middle map is the medium map with hidden terrain, and the right map is the largest map with hidden terrain. The green grids on the maps represent hidden terrain where the operators' observed distance is reduced.}
   \label{map3}
\end{figure*}

\textbf{Map and Rule}.
To enable greater freedom of movement, WGC uses a hexagonal map, similar to that in wargames. Certain hexagons on the map possess a hidden terrain property, making it difficult for enemies to observe operators located within them due to the reduced observed distance. This feature allows for the partial observability to be reflected even when operators are fighting at a close distance. WGC offers three shared maps for all sub-environments, comprising a small-sized map with no terrain, a medium-sized map with hidden terrain, and a large-sized map with hidden terrain (as shown in Figure \ref{map3}). Additionally, WGC incorporates various distinct rules, including rules for movement, visibility, attack, and so on. For example, "Move", which is a critical action in the game, different operators may require varying time steps to move to an adjacent hexagon, which is a crucial factor that makes WGC an asynchronous game. For attack, When an agent attacks another agent, it causes damage based on a random seed, making WGC a random environment.


\begin{table*}[htbp]
\begin{center}
\caption{Operators attributes.}
\label{att}
\begin{tabular}{l l}
\hline
Attributes     &   Description   \\
\hline
color & 0 (red) or 1 (blue)  \\
id & decimal number  \\
type & 0 for tank, 1 for chariot, and 2 for infantry  \\
blood & current blood  \\
position & current position  \\
speed & move speed  \\
observed distance & max distance an operator can be observed  \\
attacked distance & max distance  an operator can be attacked  \\
attack damage(when attack a tank or chariot)  & attack capability  \\
probability of causing damage (when attack a tank or chariot)  &  attack random seed  \\
attack damage(when attack a infantry)  & attack capability  \\
probability of causing damage(when attack a infantry)  & attack random seed \\
shoot cool-down time & cool-down time after shoot  \\
shoot preparation time & shoot preparation time after move   \\
move time & move cost \\
stop time & stop cost \\
\hline
\end{tabular}
\end{center}
\end{table*}

\begin{table*}[h]
\begin{center}
\caption{Operators attributes template.}
\label{tab:tank}
\begin{tabular}{l l l l}
\hline
Attributes     &Tank   &chariot &infantry   \\
\hline
blood & 10 &8 &7\\
speed & 1 &1 &0.2\\
observed distance & 10 &10 &5\\
attacked distance & 7 &7 &3\\
attack damage  & 1.2 &1.5 &0.8 \\
probability of causing damage  &  0.8 &0.6 &0.6  \\
attack damage  & 0.6 &0.8 &0.8  \\
probability of causing damage  & 0.4 &0.6 &0.6 \\
shoot cool-down time & 0  &1 &1 \\
shoot preparation time & 0  &2 &2   \\
\hline
\end{tabular}
\end{center}
\end{table*}

\textbf{Built-in Opponent Bot}.
WGC provides two built-in bots that enable RL agents or human players to compete against those rule-based bots with diverse strategies:
\begin{itemize}
\item KAI0: a rush AI with a high priority on attacking and a strategy of heading directly to the center of the battlefield.
\item KAI1: an AI that uses a commonly employed strategy in wargames, leveraging hiding tactics to ambush enemies near special terrain.
\end{itemize}

\subsubsection {Scenarios And Modification} 
In WGC, each sub-environment contains three scenarios for studying problems of different scales, and easy modification is also a key feature of WGC.

\textbf{Scenarios}.
A Scenario is a combination of operator, map, and rule. We have designed a unique set of operator parameters for each environment to implement its corresponding research topic.  Each scenario comprises one map and one set of operator parameters, resulting in three scenarios for each environment. 

\begin{figure}[!h]
   \begin{center}
   \includegraphics[width=0.45\textwidth]{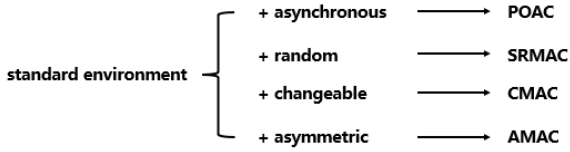}
   \end{center}
   \caption{Relation between five sub-environments}
   \label{guanxi2}
\end{figure}

\textbf{Sub-Environment Modification}.
The development of the five sub-environments is all based on these shared elements. By modifying these elements, it is easy to obtain sub-environments with different functionalities, WGC offers a set of common interfaces that allow users to easily modify the environment codes and create different functional sub-environments.
In addition to directly modifying the codes, WGC also allows users to modify scenario settings by editing configuration files. This makes it easy to fine-tune the environment to meet specific needs. For instance, by editing operator attributes in JSON files such as "speed", the POCA environment can be transformed into a synchronous cooperation game. Users can also create their own maps by editing map data in HDF5 files.

\subsubsection {Shared Elements For MARL}
To facilitate the training of multi-agent reinforcement learning, we have designed a set of standard features, actions, and rewards for all sub-environments.

\textbf{Observation and State}.
At each time step, every agent in WGC receives local observation within its field of view. Specifically, one agent can observe other agents only if they are both alive and located within the sight range. This limited sight range renders the environment partially observable from the standpoint of each agent.
The feature vector observed by each agent contains critical attribute information, including time and attack information. Furthermore, the global state, which is only available to agents during centralized training, provides perfect information about all operators on the map. All features are normalized by their maximum values.
It is worth noting that we provide map features for more comprehensive information with an open interface.

\textbf{Action Space}.
In WGC, the agents can take move, shoot, and stop actions. The move action allows agents to move in six directions, one hexagon per time. However, each move action requires a certain number of time steps to complete, which is different for each operator type. The shoot action allows agents to attack a specific enemy which is in sight, identified by their ID, within their shooting range. However, before taking a shoot action, the agent must first spend a certain amount of time in preparation. Once the preparation time is finished, the shoot action will occur. Finally, the stop action allows agents to perform no action in that time step.

\textbf{Reward}.
The overall goal is to maximize the win rate for each battle.
The default reward setting is computing the disparity between the amount of health lost by allied forces and that of the enemy's.

\subsection{Sub-Environments}
In WGC, the standard environment serves as a benchmark sub-environment upon which other sub-environments can be constructed by introducing different settings. When the standard environment is equipped with asynchronous functionality, it becomes the POAC sub-environment. When strong randomness is added, it becomes the SRMAC sub-environment. When polymerization and depolymerization actions are introduced, it becomes the CMAC sub-environment. When asymmetric operator setting is added, it becomes the AMAC sub-environment. The relationship between the five sub-environments is illustrated in Figure \ref{guanxi2}. The existence of the standard environment allows us to observe how introducing different new features to partially observable multi-agent cooperation environments will affect algorithm performance. 

\subsubsection{Standard Environment}
The standard environment is primarily designed for  partially observable fully cooperative multi-agent tasks, which is an attractive research area. Such problems are relevant to a large number of real-world systems and are more amenable to evaluation than general-sum problems. Consequently, there have been many studies on this task \cite{smac}. Generally, the cooperative multi-agent task is described as Dec-POMDPs \cite{oliehoek2016concise}. Formally, a Dec-POMDP $G$ is given by a tuple $G=<S, U, P, r, Z, O, n, \gamma>$, and $s \in \mathrm{S}$ is the true state of the environment. At each time step, each agent $a \in A \equiv\{1, \ldots, n\}$ chooses an action $u^{a} \in U$ and forms a joint action $\mathbf{u} \in \mathbf{U} \equiv U^{n}$, which causes a transition of the environment according to the state transition function $P\left(s^{\prime} \mid s, \mathbf{u}\right): S \times \mathbf{U} \times S \rightarrow[0,1]$.

\textbf{Action Space}. 
For each operator, the action space includes six movement directions $\mathbf{move}[direction]^{6}$, shoot enemies $\mathbf{shoot}[enemy\_id]^{enemy\_num}$, and stop.

\textbf{Scenarios}.
In the standard environment, each team consists of three operators: a chariot, a tank, and an infantry. The settings are minimal, with all operators moving instantly and no cool-down or preparation time for shooting. This makes the standard environment relatively simple. The operator parameters of the standard environment are listed in Table \ref{tab:standard}.

\textbf{Environment modification}. 
The standard environment can be easily fine-tuned since it includes many common settings, there is still a lot of potential for improvement. We leave these improvements to the users.

\begin{table*}[htbp]
\begin{center}
\caption{Standard sub-environment operators attributes setting. The red side and blue side  have the same setting.}
\label{tab:standard}
\begin{tabular}{l l l l}
\hline
Attributes     &Tank   &chariot &infantry   \\
\hline
blood & 10 &8 &7\\
speed & 1 &1 &1\\
observed distance & 10 &10 &5\\
attacked distance & 7 &7 &3\\
attack damage  & 1.2 &1.5 &0.8 \\
probability of causing damage  &  0.8 &0.7 &0.7  \\
attack damage  & 0.6 &0.8 &0.8  \\
probability of causing damage  & 0.6 &0.6 &0.6 \\
shoot cool-down time & 0  &0 &0 \\
shoot preparation time & 0  &0 &0   \\
\hline
\end{tabular}
\end{center}
\end{table*}

\begin{table*}[htbp]
\begin{center}
\caption{CMAC sub-environment operators attributes setting. The red side and blue side  have the same setting.}
\label{tab:cmac}
\begin{tabular}{l l l l}
\hline
Attributes     &Tank   &chariot &infantry   \\
\hline
blood & 10 &8 &7\\
speed & 1 &1 &1\\
observed distance & 10 &10 &5\\
attacked distance & 7 &7 &3\\
attack damage  & 1.2 &1.5 &0.8 \\
probability of causing damage  &  0.8 &0.7 &0.6  \\
attack damage  & 0.6 &0.8 &0.8  \\
probability of causing damage  & 0.6 &0.6 &0.6 \\
shoot cool-down time & 0  &0 &0 \\
shoot preparation time & 0  &0 &0   \\
attack reduce coefficient & 0.8  &0.8 &0.8   \\
\hline
\end{tabular}
\end{center}
\end{table*}

\begin{table*}[htbp]
\begin{center}
\caption{AMAC sub-environment operators attributes setting. The red side has 2 tanks, 2 chariots and 1 infantry,the blue side has 1 tanks, 1 chariots and 1 infantry, the red side and the blue side are different in attributes setting.}
\label{tab:amacr}
\begin{tabular}{l l l l l l l}
\hline
Attributes     &Tank\_red(2)   &chariot\_red(2) &infantry\_red(1)  &Tank\_blue(1)   &chariot\_blue(1) &infantry\_blue(1)   \\
\hline
blood & 10 &8 &7 & 10 &8 &7\\
speed & 1 &1 &1 & 1 &1 &1\\
observed distance & 10 &10 &5 & 10 &10 &5\\
attacked distance & 7 &7 &3 & 7 &7 &3\\
attack damage  & 1.5 &1.5 &0.8 & 1.2 &1.5 &0.8 \\
probability of causing damage   &  0.8 &0.7 &0.7 &  0.8 &0.6 &0.6  \\
attack damage  & 0.8 &0.8 &0.8 & 0.6 &0.8 &0.8  \\
probability of causing damage  & 0.6 &0.6 &0.6 & 0.6 &0.6 &0.6 \\
shoot cool-down time & 0  &0 &0 & 0  &0 &0 \\
shoot preparation time & 0  &0 &0  & 0  &0 &0 \\
\hline
\end{tabular}
\end{center}
\end{table*}

\subsubsection{POAC}
POAC is designed for the partially observable asynchronous multi-agent cooperation task.
In most multi-agent benchmarks, synchrony between actions is an implicit assumption. To the best of our knowledge, POAC is the first partially observable asynchronous multi-agent cooperation challenge for the MARL community.

It is called synchronous if there is a global clock and agents move in lockstep and \( \operatorname{step}_{j}^{a_{i}} \pm \) in the system corresponds to a tick of the clock.
However, in an asynchronous cooperation system, there is no global clock, and the agents in the system can run at arbitrary rates relative to each other \cite{halpern2007computer}.
Because of this, Dec-POMDPs should be modified for asynchronous tasks.
Specifically, at each time step (suppose a global clock exists), each agent chooses an action $u^{a} \in A$ forming a joint action $\mathbf{u}^{val} \in \mathbf{U}^{val} $, where $\mathbf{u}^{val}$ is actions of agents that can perform actions. This will cause a transition of the environment $P\left(s^{\prime} \mid s, \mathbf{u}^{val}\right): S \times \mathbf{U}^{val} \times S \rightarrow[0,1]$.
In POAC, heterogeneous agents have their own action spaces and executed cycle, which are different from each other.

\begin{figure*}
   \begin{center}
   \includegraphics[width=0.8\textwidth]{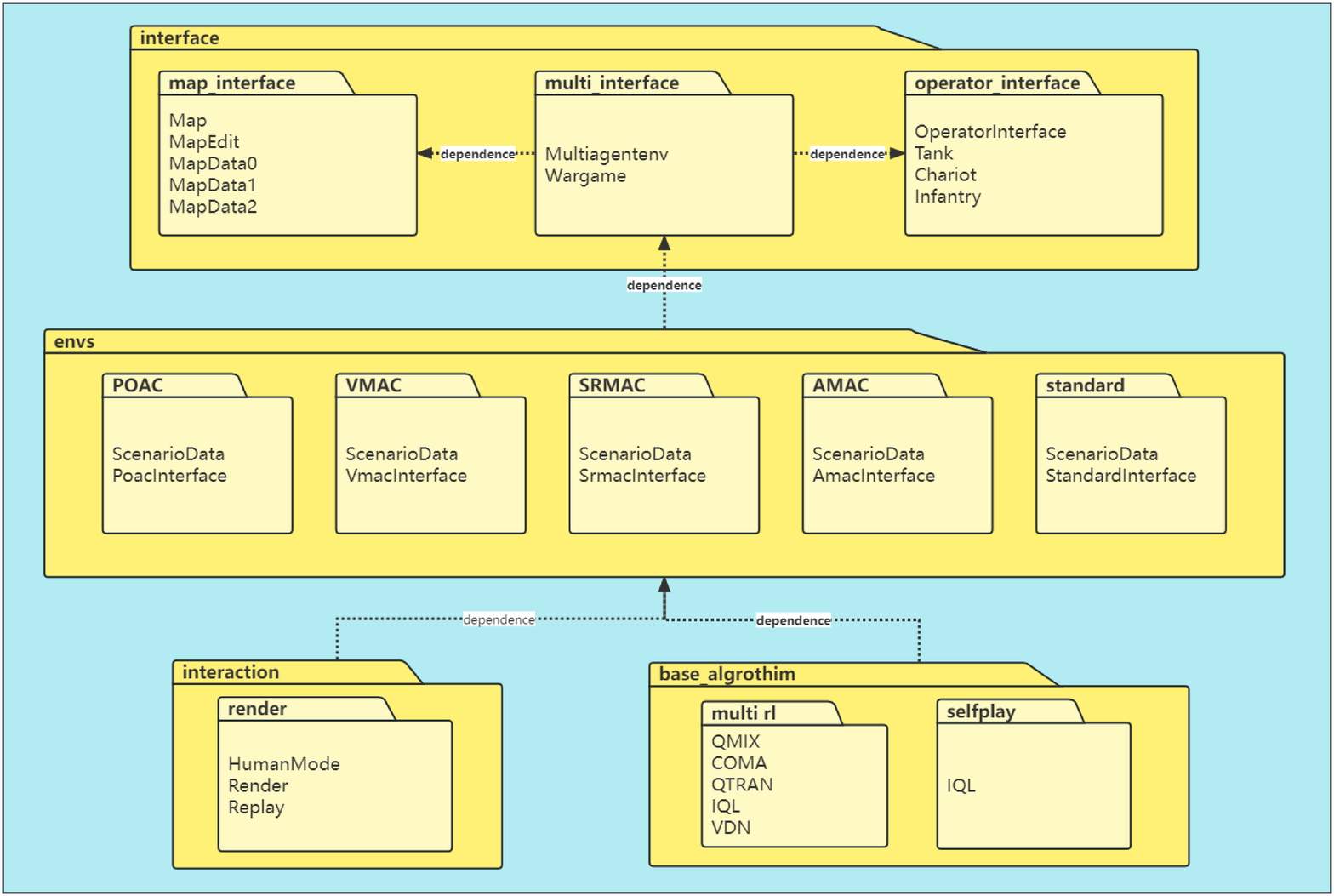}
   \end{center}
   \caption{Framework of WGC}
   \label{framework}
\end{figure*}

\textbf{Action Space}. 
For each operator, the action space includes six movement directions $\mathbf{move}[direction]^{6}$, shoot enemies $\mathbf{shoot}[enemy\_id]^{enemy\_num}$, and stop.

\textbf{Scenarios}
In POAC, each team consists of three operators: a chariot, a tank, and an infantry. We have configured the chariot and tank to move five times faster than the infantry. Additionally, we have set a shoot preparation time for the chariot and infantry operators, but not for the tank operators.

\textbf{Environment modification}. 
Users can modify the attributes of the operators in POAC, such as adjusting the move time, to control the execution cycle of the asynchronous action space.

\subsubsection{CMAC}
CMAC is designed for changeable agents multi-agent cooperation task. Inspired by the formation mode in wargame, where multiple soldiers can form a formation and be disbanded into individual soldiers, we have designed a special action space for CMAC. This action space allows agents to be decomposed into multiple smaller agents, which can then be reassembled into a single agent. This setting is unique compared to most multi-agent environments and may require careful consideration of model transfer learning.

\textbf{Action Space}.
For each operator, the base action space includes six movement directions $\mathbf{move}[direction]^{6}$, shoot 
enemies $\mathbf{shoot}[enemy\_id]^{enemy\_num}$, and stop. In addition to the base action space, we have introduced two unique actions: polymerization and depolymerization. The depolymerization action allows an agent to break down into either three smaller agents or one medium-sized agent and one small agent $\mathbf{depolymerization}[num]^{2}$. Conversely, the polymerization action enables two decomposed agents to reassemble into one larger agent $\mathbf{polymerization}[ally\_id]^{2}$. 

\textbf{Scenarios}.
In CMAC, each team includes three operators: a chariot, a tank, and an infantry. The majority of settings in CMAC are similar to the standard environment. However, when an agent performs the depolymerization action, a new ID is generated for each new agent. Additionally, when an agent decomposes into multiple small agents, the attack ability of each small agent decreases, when perform polymerization action, small agents reassemble into one larger agent, the attack ability is restored. The operator parameters for CMAC are listed in table \ref{tab:cmac}.

\textbf{Environment modification}. 
The CMAC environment comes with a set of predefined settings, with only limited options for customization. For example, users can modify is the degree of attack ability decrease when an agent is decomposed into smaller agents.

\subsubsection{AMAC}
AMAC is designed for asymmetric multi-agent cooperation task. Asymmetry refers to the unbalanced levels of ability between different sides in games. Most games typically ensure that different sides in the game have relatively balanced abilities. For instance, in the StarCraft game, there are three races, each with completely different technology trees and operator types, but the three races are roughly balanced in terms of ability. However, wargame is an unbalanced game, with asymmetry mainly reflected in the gap in the number and capabilities of the two sides. In AMAC, we reflect this asymmetry by setting the operators' number and attack capabilities of the two sides accordingly.

\textbf{Action Space}. 
For each operator in AMAC, the action space includes six movement directions $\mathbf{move}[direction]^{6}$, shoot enemies $\mathbf{shoot}[enemy\_id]^{enemy\_num}$, and stop.

\textbf{Scenarios}.
In AMAC, the blue team includes five operators with reinforced attack ability, two chariots, two tanks, and one infantry. The red team, on the other hand, includes three operators with normal attack ability, a chariot, a tank, and an infantry. The operator parameters of AMAC are listed in tables \ref{tab:amacr}.

\textbf{Environment modification}. 
In AMAC, users have the flexibility to increase the number of operators or adjust their attack abilities to further customize the environment. This allows for experimentation with different levels of asymmetry between the two sides and can provide insights into how varying levels of imbalance impact cooperative behavior.

\subsubsection{SRMAC}
SRMAC is designed for strongly stochastic and high-risk multi-agent cooperation task.
The high level of risk is due to the randomness of judgment in the game. In games like StarCraft, although the damage value is fixed, it is still affected by small stochastic factors. However, in wargame, all attacks are affected by stochastic factors, which greatly influences the game outcome. Furthermore, unlike other real-time strategy games, in wargame, once the troops are eliminated, they will not be regenerated, resulting in extremely high risks. To reflect this randomness, SRMAC is designed with multi-conditional random numbers.

\textbf{Action Space}. 
For each operator in AMAC, the action space includes six movement directions $\mathbf{move}[direction]^{6}$, shoot enemies $\mathbf{shoot}[enemy\_id]^{enemy\_num}$, and stop.

\textbf{Scenarios}.
In SRMAC, each team includes three operators: a chariot, a tank, and an infantry. The majority of the settings are similar to the standard environment. After attacking an enemy, there is a probability that the enemy will lose all of their health and disappear. Additionally, there is a probability that the attack will not work. Most of the time, the attack will cause a damage value that is affected by the distance and the health of the enemy.

\textbf{Environment modification}. 
In addition to the modifications that can be made in the standard environment, SRMAC introduces further modifications to the attack interface to add more randomness. This allows users to customize the level of stochasticity in the game, reflecting the high risk nature of the game.

\subsection{WG Challenge Engine Framework}

In the WG Challenge engine, there are four main modules as shown in Figure \ref{framework}:

\begin{enumerate}[]
\item The Interface module, which is the core component of the WGC engine, includes common interface functions and three default map data that support five sub-environments.
\item The Envs module, which includes five independent sub-environment projects, each with a default scenario setting file.
\item The Algorithm module, which provides five multi-agent reinforcement learning algorithms, including Qmix, Qtran, VDN, IQL, COMA, and a self-play interface.
\item The Interaction module, which provides a human mode where humans can play against other humans or AI models, and a replay mode for reviewing each game.
\end{enumerate}

We offer the environment interface with a combination of gym interface and Pymarl interface, since the gym interface does not consider multi-agent situations, we set the Pymarl\cite{smac} interface to the upper layer of the gym interface. We consider that this combination has the potential to provide standardized interfaces for different multi-agent environments.
WGC offers two additional control modes. The first one is the self-Play mode, and the second mode is the human mode, where human players can use WGC to battle against the built-in bots, RL agents, and other human players.

\section{Experiments}
In this section, we will present experimental results of WGC, which are based on all scenarios of five sub-environments, including the win rate results of the competition between built-in knowledge AI and built-in knowledge AI, as well as the win rate results of the competition between built-in knowledge AIs and QMIX agents.The win rate is an important metric as it stands for the game level of the agents.

\begin{figure*}[h!]
\centering
\subfigure{  
\includegraphics[width=3.4cm]{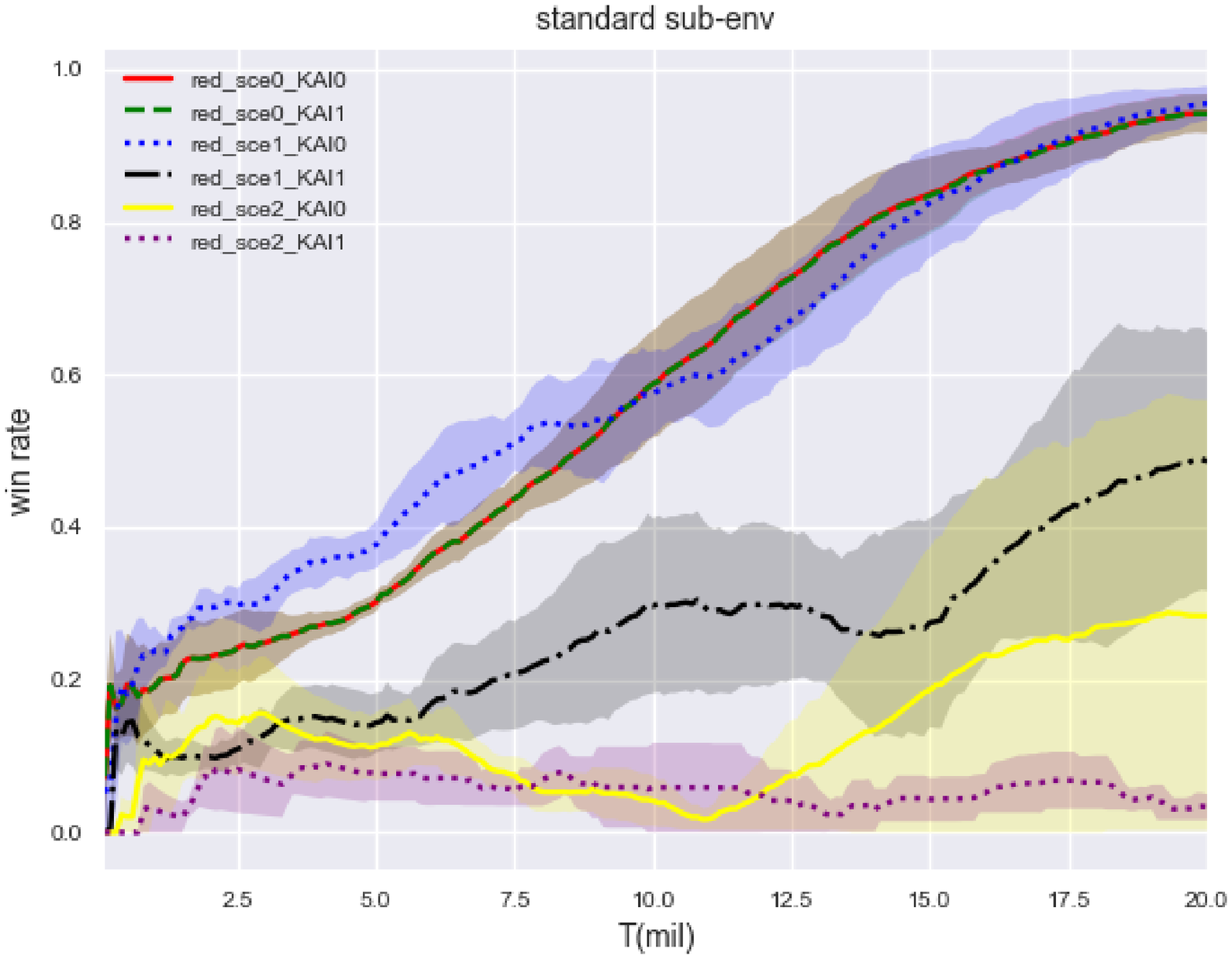} 
\includegraphics[width=3.4cm]{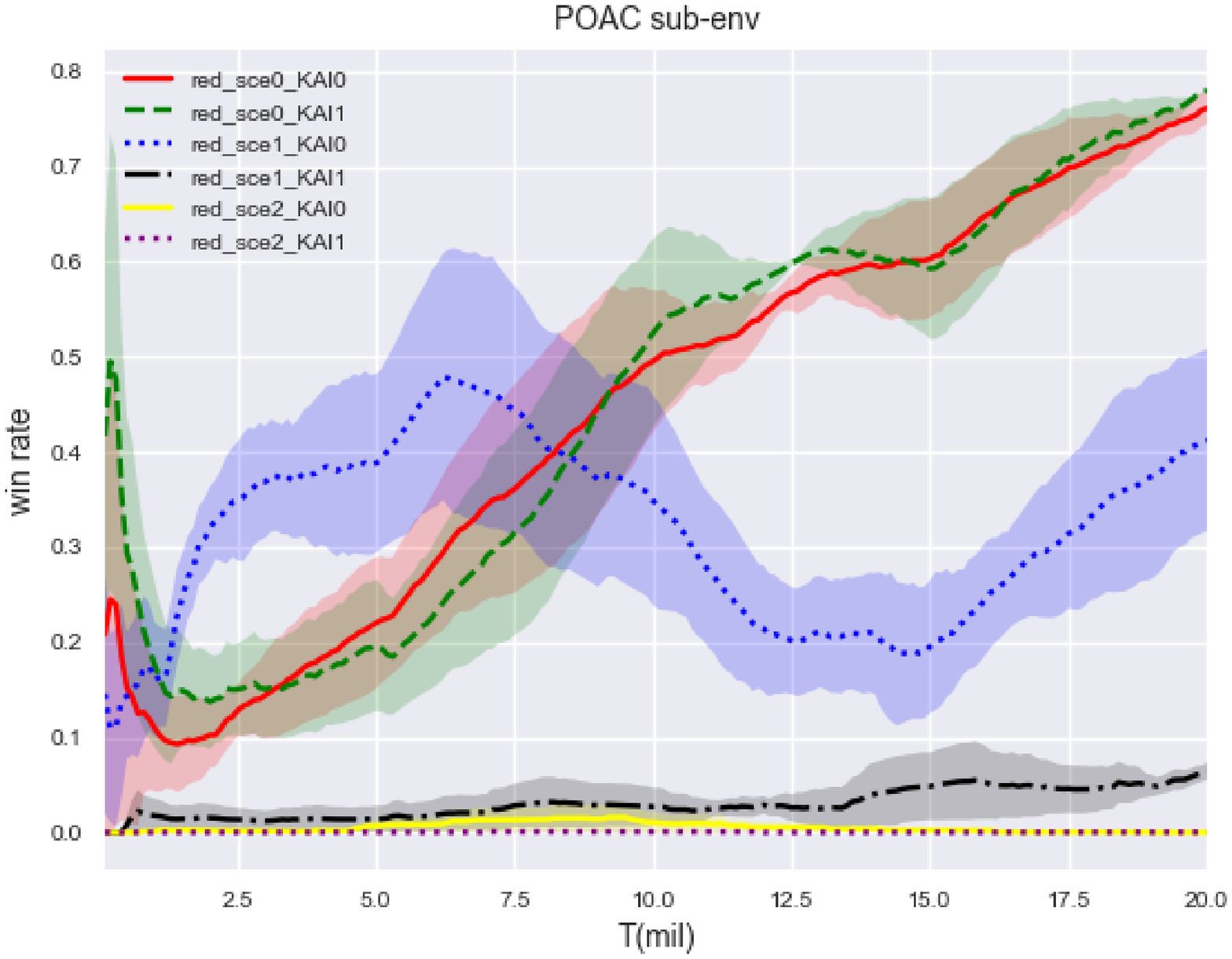}
\includegraphics[width=3.4cm]{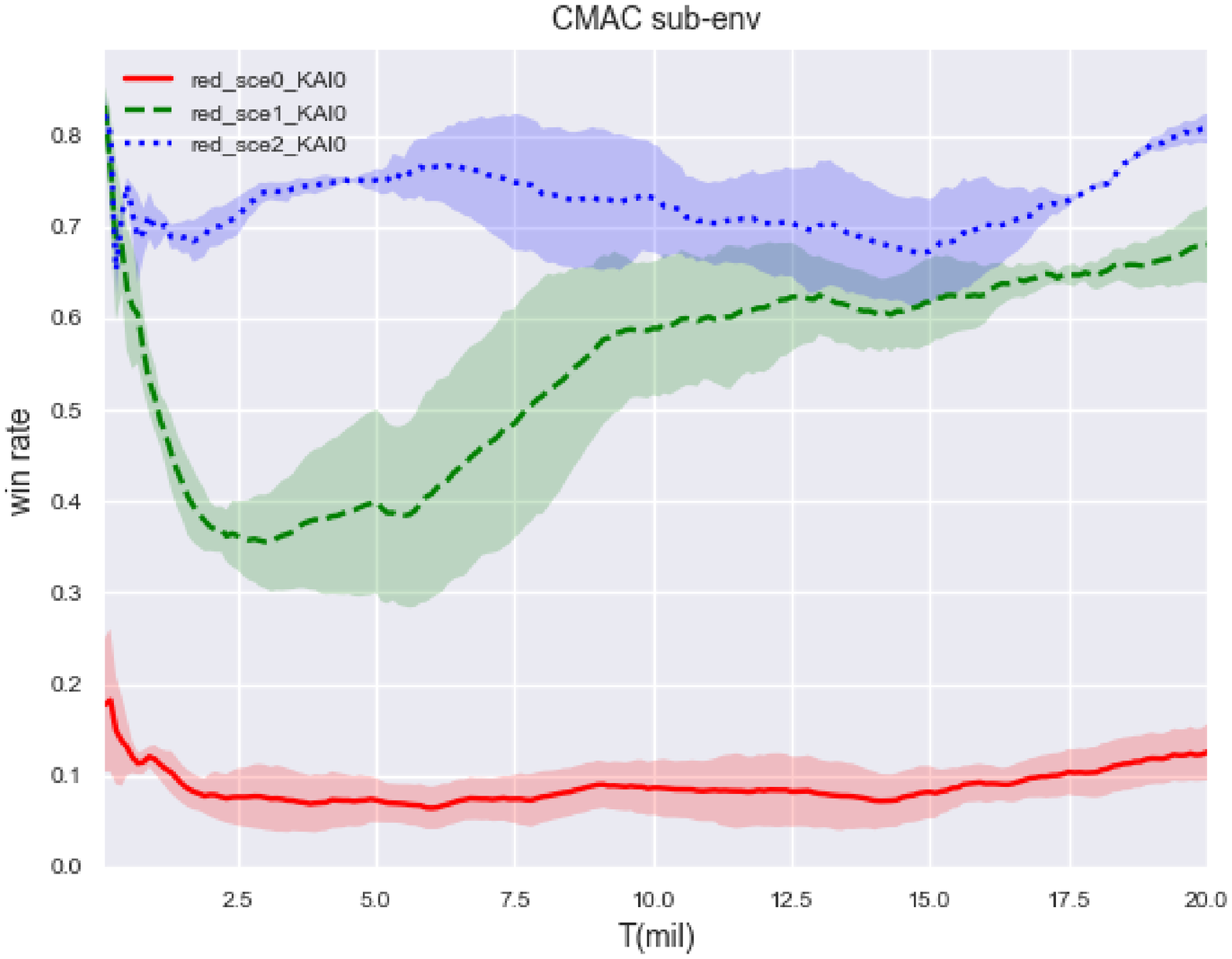}
\includegraphics[width=3.4cm]{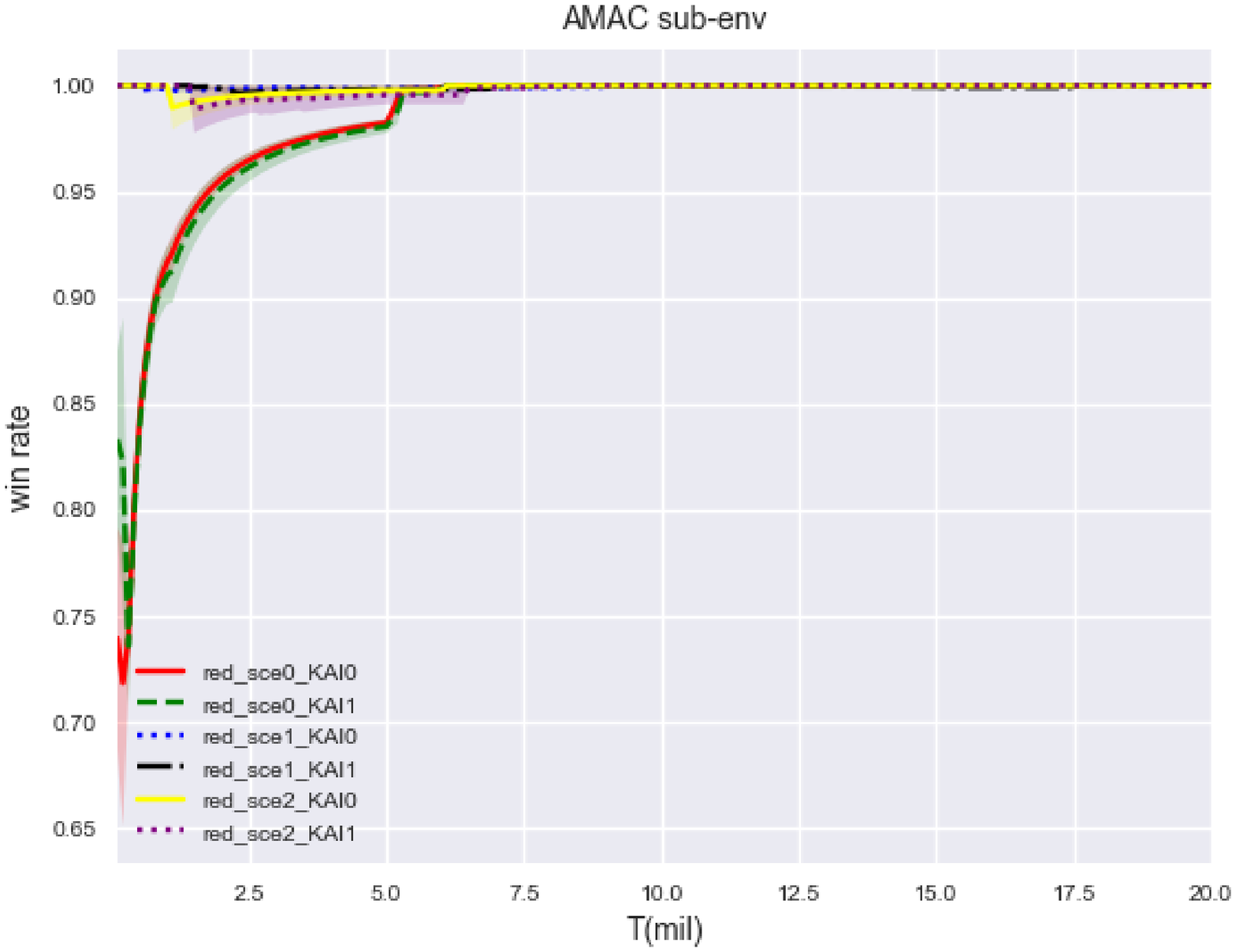}
\includegraphics[width=3.4cm]{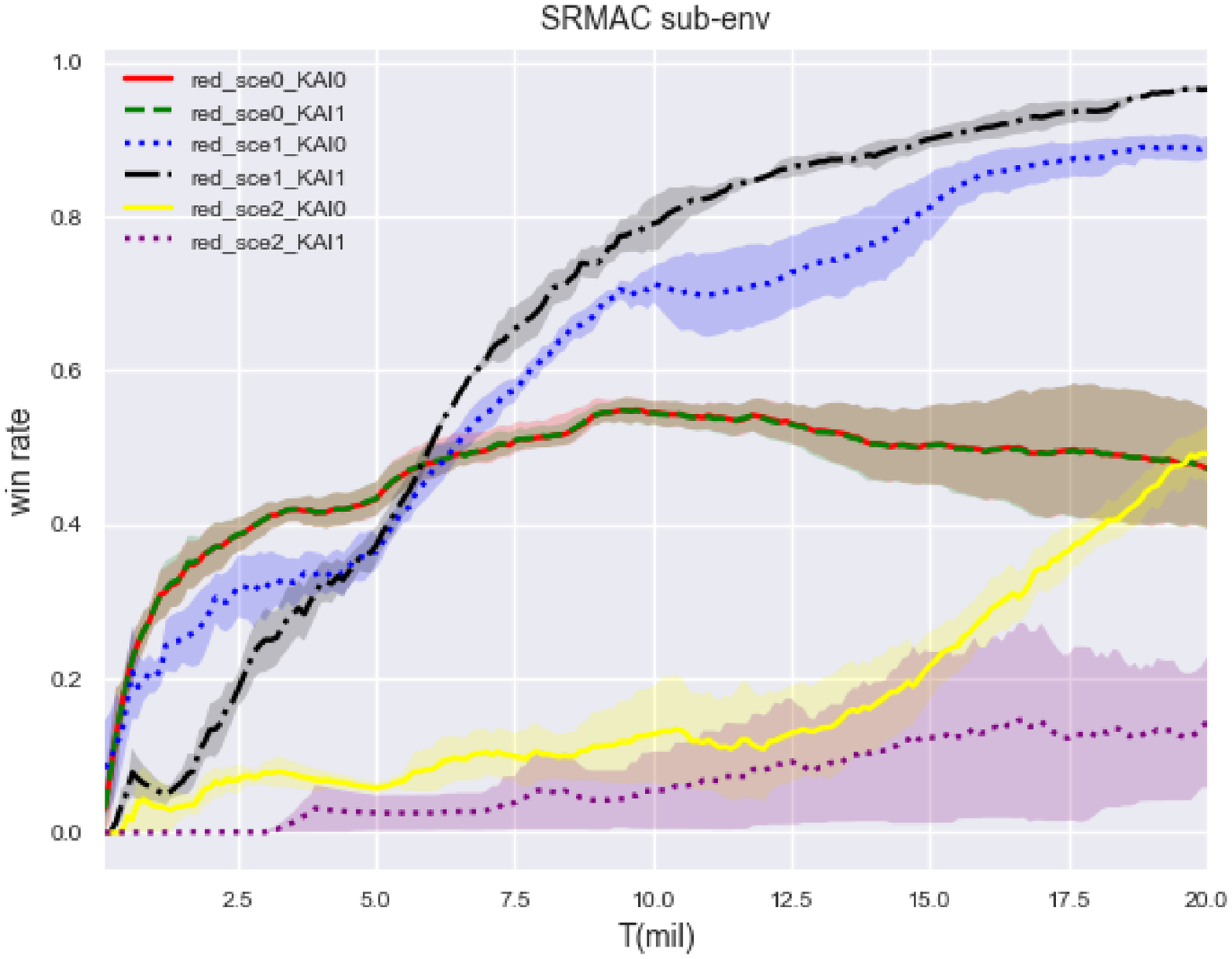}
}
\quad
\subfigure{ 
\includegraphics[width=3.4cm]{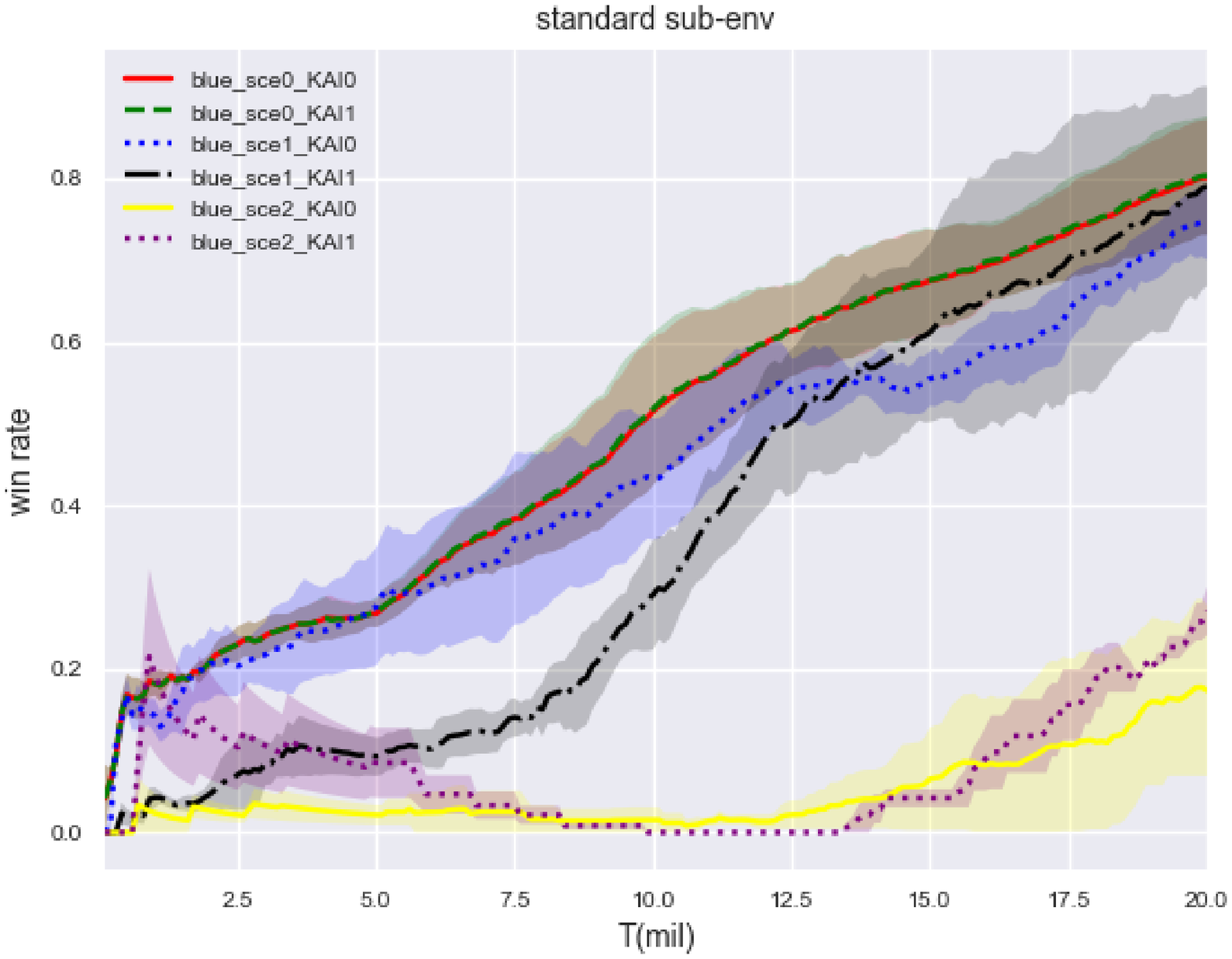}
\includegraphics[width=3.4cm]{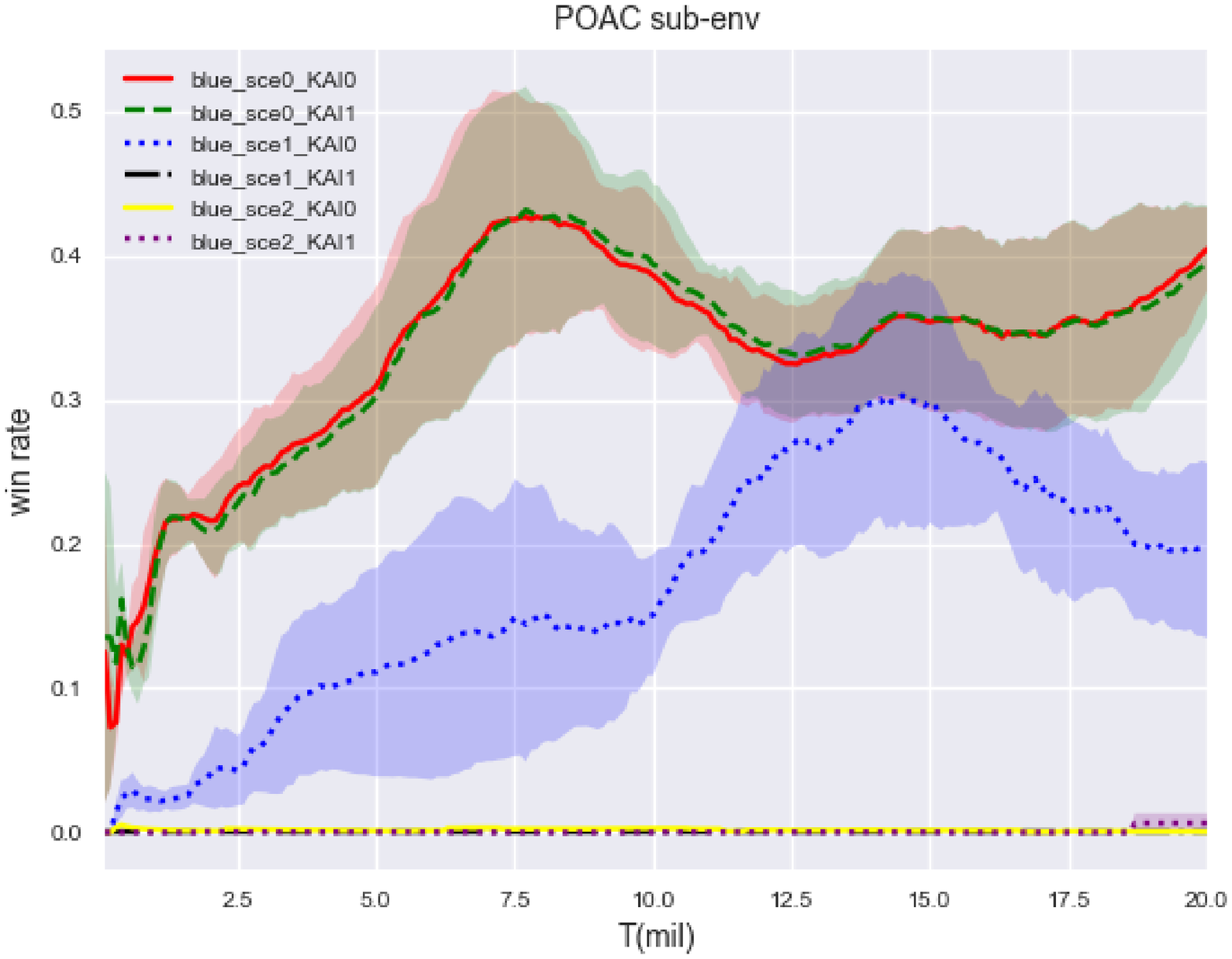}
\includegraphics[width=3.4cm]{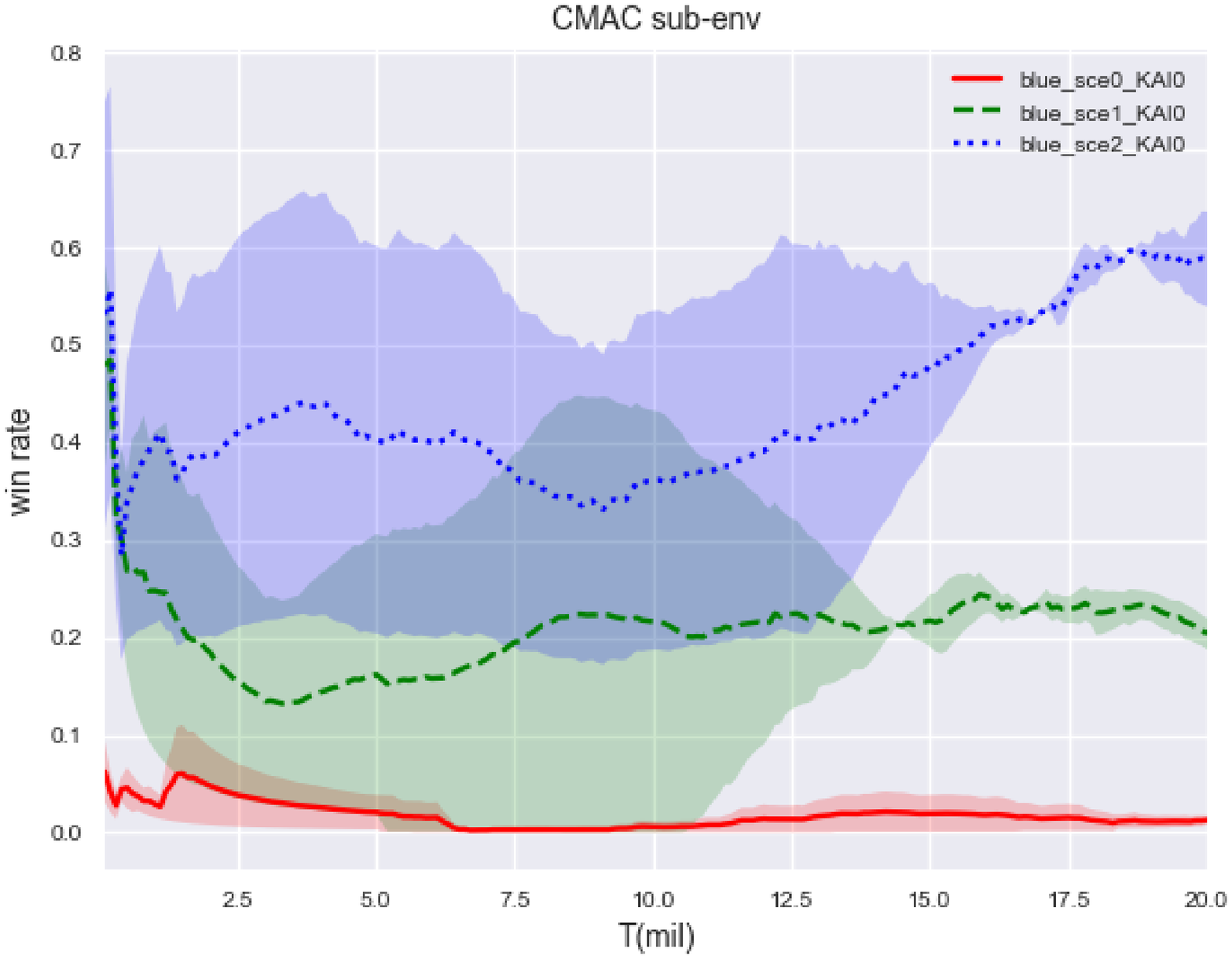}
\includegraphics[width=3.4cm]{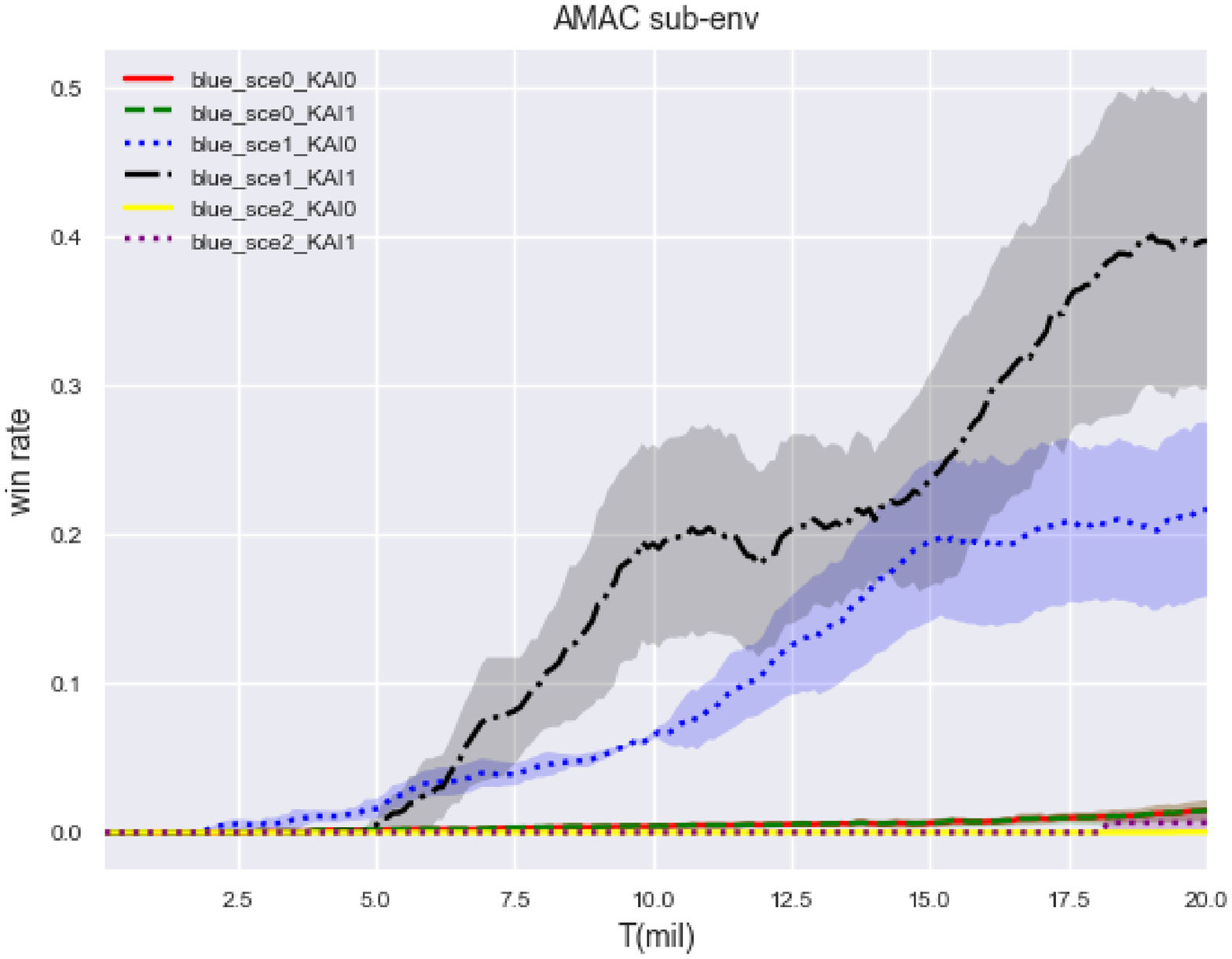}
\includegraphics[width=3.4cm]{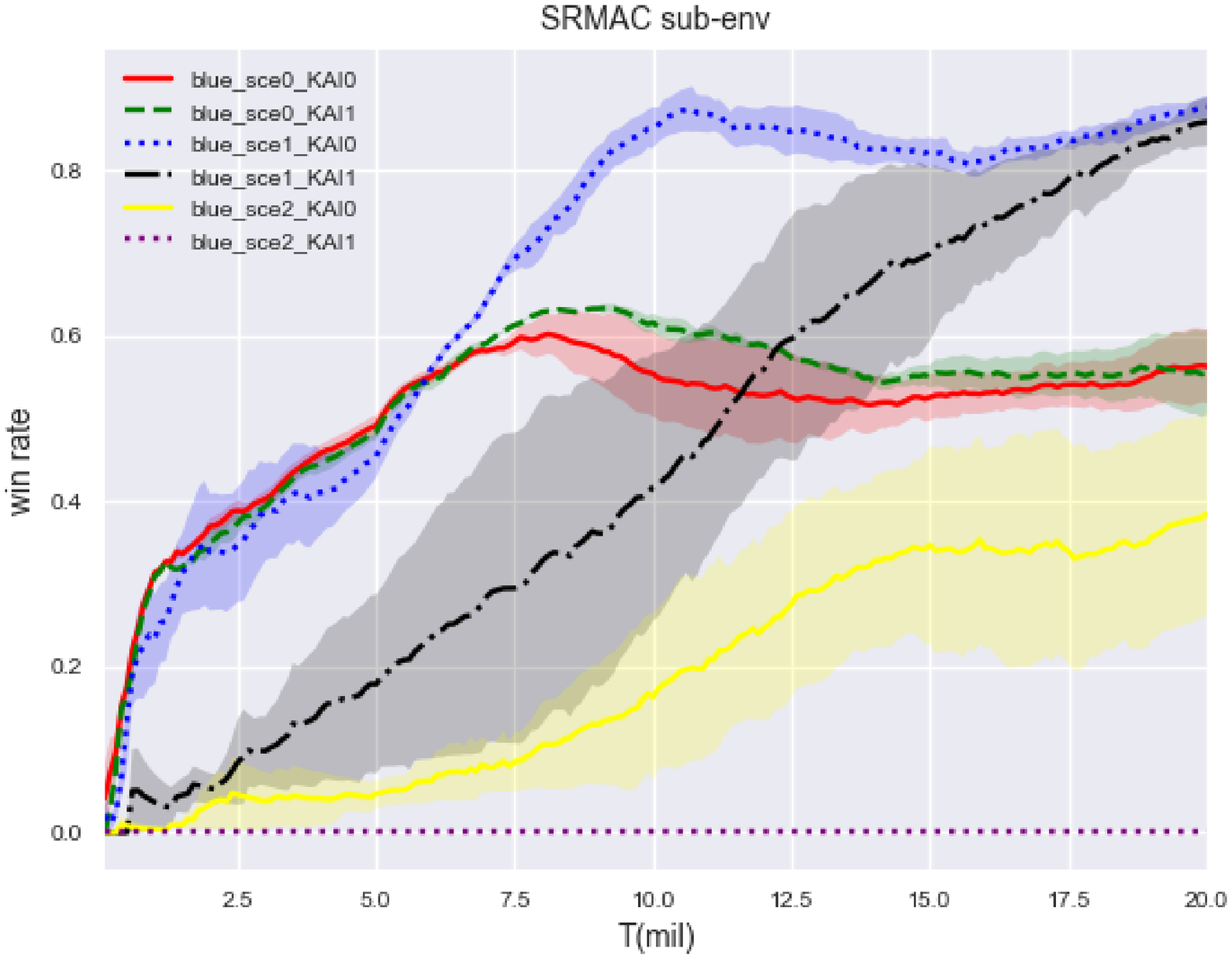}
}
\caption{QMIX experiment results on all scenarios of all sub-environments. red\_sce0\_kai0 means QMIX agent as red side compete against KAI0 in scenario 0. }    
\label{resultexe}       
\end{figure*}

\begin{figure*}[h!]
   \begin{center}
   \includegraphics[width=0.6\textwidth]{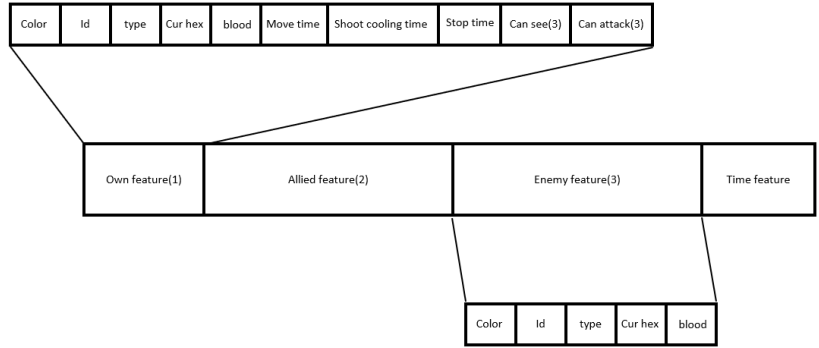}
   \end{center}
   \caption{Feature space.The feature space of each operator is composed of its own feature, ally feature, enemy feature, and time feature. Some enemy feature items cannot be obtained as they are confidential information.}
   \label{tz}
\end{figure*}

\begin{figure}
   \begin{center}
   \includegraphics[width=0.45\textwidth]{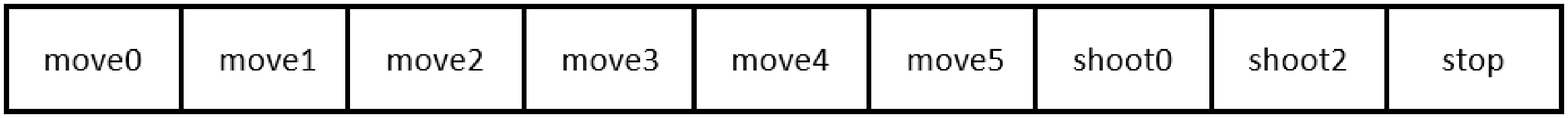}
   \end{center}
   \caption{Action space.As an example of standard sub-environment action space, the action space is a 9-dimensional vector respectively represent move action, shoot action, and stop action.}
   \label{ca}
\end{figure}

\subsection{Experimental Settings}
For environment settings, We use the default settings for all scenarios, as introduced in the previous section. We set each game with 600 time steps, and when time is reached or one side in red and blue is all dead, the game is over.
For knowledge AI experiment, we design two different built-in bots that use distinct types of strategies. KAI0 built-in bots can be applied in all scenarios of all sub-environments and served as red or blue. KAI1 built-in bots can be applied in most sub-environments except CMAC (only KAI0 can be applied in CMAC) and served as red or blue. 

For MARL network settings, the architecture of each agent network is a DRQN with one recurrent layer. Specifically, the network consists of a GRU with a 64-dimensional hidden state and a fully-connected layer before and after. For feature and action setting, details of feature encoding and action space encoding can be found in Figures \ref{tz} and \ref{ca}, respectively. The feature matrix of each operator consists of four parts as shown in the figure. The attribute information of the operator itself is encoded at the forefront, and the attribute information of the teammates is encoded in sequence according to their IDs. These two parts of information include the color, id, type, current hex, blood, move time, shoot cooling time, stop time, observation information and attack information of each operator. The enemy operator information is encoded in the third part according to their IDs, and their attributes information only includes the color, id, type, current hex and blood, other information is not directly accessible, which is consistent with the setting of wargame. The final part of the features is the game time step information. The action space of different sub-environments is different, but the encoding organization of the action space is similar across different sub-environments. Taking POAC as an example, actions are encoded in sequence as shown in the figure \ref{ca}. 

We follow QMIX experimental settings of Pymarl\cite{smac} except that we use an empty action when no executed actions can be obtained for an agent in POAC. We set up action mask, which can filter out invalid actions during training due to the asynchronous nature of actions. For CMAC, we set QMIX input with the shape of the maximum number of agents in CMAC.
For evaluation, we pause the training after every 10,000 time steps and run 32 test episodes, during which the agents perform action selection greedily in a decentralized fashion. The percentage of winning episodes is reported, where a win is recognized when the agents' blood is larger than the enemies' within 600 time steps. Each independent run of the complete training experiment takes between 36 to 40 hours, depending on the exact scenario, using Nvidia Geforce RTX 2080 Ti graphics cards.

\begin{table}[h!]
\begin{center}
\caption{Results of built-in bots compete against each other in all scenarios of all sub-environments, and the number xxx/xxx means win rate/average time steps used.}
\label{result}

\begin{tabular}{l l l l}
stand\_sce 0 & KAI0-blue & KAI1-blue \\ 
\hline
KAI0-red & 0.450/114.7 & 0.500/278.3\\
KAI1-red & 0.510/109.1 & 0.470/110.2 \\
\hline
\end{tabular}

\vspace{1em}

\begin{tabular}{l l l l}
stand\_sce 1 & KAI0-blue &  KAI1-blue   \\
\hline
KAI0-red & 0.460/144.1 & 0.440/338.6  \\
KAI1-red & 0.660/271.7 & 0.350/599.0 \\
\hline
\end{tabular}

\vspace{1em}

\begin{tabular}{l l l l}
stand\_sce 2 & KAI0-blue &  KAI1-blue  \\
\hline
KAI0-red & 0.420/187.2 & 0.500/106.2 \\
KAI1-red & 0.340/339.0 & 0.050/599.0 \\
\hline

\end{tabular}

\vspace{1em}

\begin{tabular}{l l l l}
poac\_sce 0 & KAI0-blue & KAI1-blue \\ 
\hline
KAI0-red & 0.490/318.3 & 0.470/346.1\\
KAI1-red & 0.480/316.3 & 0.430/323.2\\
\hline
\end{tabular}

\vspace{1em}

\begin{tabular}{l l l l}
poac\_sce 1 & KAI0-blue &  KAI1-blue   \\
\hline
KAI0-red & 0.470/433.2 & 0.200/530.2  \\
KAI1-red & 0.660/521.1 & 0.420/594.7 \\
\hline
\end{tabular}
\vspace{1em}

\begin{tabular}{l l l l}
poac\_sce 2 & KAI0-blue &  KAI1-blue  \\
\hline
KAI0-red & 0.300/527.7 & 0.100/574.4 \\
KAI1-red & 0.660/589.3 & 0.020/599.0 \\
\hline
\end{tabular}

\vspace{1em}

\begin{tabular}{l l l l}
cmac\_sce 0 & KAI0-blue\\
\hline
KAI0-red & 0.700/219.9\\
\hline
\end{tabular}

\vspace{1em}

\begin{tabular}{l l l l}
cmac\_sce 1 & KAI0-blue   \\
\hline
KAI0-red & 0.590/340.3  \\
\hline
\end{tabular}

\vspace{1em}

\begin{tabular}{l l l l}
cmac\_sce 2 & KAI0-blue  \\
\hline
KAI0-red & 0.680/384.2\\
\hline
\end{tabular}

\vspace{1em}

\begin{tabular}{l l l l}
amac\_sce 0 & KAI0-blue & KAI1-blue \\ 
\hline
KAI0-red & 1.000/85.0 & 1.000/79.5\\
KAI1-red & 1.000/80.8 & 1.000/80.3\\
\hline
\end{tabular}

\vspace{1em}

\begin{tabular}{l l l l}
amac\_sce 1 & KAI0-blue &  KAI1-blue   \\
\hline
KAI0-red & 1.000/118.1 & 1.000/178.1  \\
KAI1-red & 1.000/515.4 & 0.980/544.7 \\
\hline
\end{tabular}

\vspace{1em}

\begin{tabular}{l l l l}
amac\_sce 2 & KAI0-blue &  KAI1-blue  \\
\hline
KAI0-red & 1.000/149.2 & 1.000/182.9 \\
KAI1-red & 1.000/599.0 & 0.960/599.0 \\
\hline
\end{tabular}

\vspace{1em}

\begin{tabular}{l l l l}
srmac\_sce 0 & KAI0-blue & KAI1-blue \\ 
\hline
KAI0-red & 0.630/112.2 & 0.710/113.8\\
KAI1-red & 0.640/102.1 & 0.660/112.2\\
\hline
\end{tabular}
\vspace{1em}

\begin{tabular}{l l l l}
srmac\_sce 1 & KAI0-blue &  KAI1-blue   \\
\hline
KAI0-red & 0.590/159.8 & 0.670/274.9  \\
KAI1-red & 0.830/257.3 & 0.380/590.1 \\
\hline
\end{tabular}
\vspace{1em}

\begin{tabular}{l l l l}
srmac\_sce 2 & KAI0-blue &  KAI1-blue  \\
\hline
KAI0-red & 0.650/200.4 & 0.630/303.0 \\
KAI1-red & 0.450/345.8 & 0.010/599.0 \\
\hline
\end{tabular}

\end{center}
\end{table}

\subsection{Results}
 Results of the competition between built-in knowledge AI can visually reflect the performance of the two AI strategies in different scenarios. The win rate matrix of built-in AI is also part of the environmental characteristics. The purpose of QMIX experiments is to provide a baseline example for future MARL studies and to demonstrate the challenge of WGC.
 
\textbf{Built-in bots vs. Built-in bots}:
Table \ref{result} presents the win rate of built-in bot combats on all scenarios of all sub-environments after 100 runs.

In scenario 0 of the standard sub-environment, there is no special terrain, the win rate for both red and blue sides is close to 0.5, and it is worth noting that KAI1's performance is close to that of KAI0 in the scenario without special terrain. In scenario 1, special terrain does not significantly increase the gap between the red and blue sides using the same strategy. When the red side uses the KAI1 strategy and the blue side uses the KAI0 strategy, the slight advantage of the KAI1 strategy is highlighted. In scenario 2, when both the red and blue sides 
use KAI1 strategy, the red side is at a significant disadvantage. On the other hand, the red side using the KAI1 strategy is slightly stronger than the blue side using the KAI0 strategy.

In the POAC sub-environment, the only change from the standard sub-environment is the addition of asynchronous operator features, which leads to significant changes in win rate matrix. In scenario 2, the characteristic of asynchrony makes the gap between the strategies and the two sides more significant than in the standard environment. 

In the CMAC sub-environment, the KAI0 of the red side has a certain strategic advantage in all three scenarios.

In the AMAC sub-environment, the asymmetrical nature makes a significant strength gap between the red and blue sides. We can see that neither of the two strategies we designed can make the weaker opponent defeat the stronger opponent. Therefore, the win rate of the red side is consistently close to 1 and is not affected by terrain and strategy changes.

In the SRMAC sub-environment, the only change from the standard sub-environment is the addition of high-randomness features, which also leads to significant fluctuations in the win rate matrix. Compared with the standard environment, the overall value of the win ratio matrix is larger.

\textbf{QMIX bots}:
 We show the QMIX experiments' results on all scenarios of all sub-environments in Figures \ref{resultexe}.
After a period of training in the standard sub-environment, the QMIX agent, as the red side, can win against KAI0 and KAI1 with a relatively high win rate in scenario 0 and against KAI0 with a relatively high win rate in scenario 1, but performs poorly in scenario 1 against KAI1 and in scenario 2 against KAI0 and KAI1. As the blue side, the QMIX agent has good performance in scenario 0 and scenario 1 but performs poorly in scenario 2.

In the POAC sub-environment, both the red and blue QMIX agents perform well in scenario 0, perform better against KAI0 than KAI1 in scenario 1, and perform poorly in scenario 2.

In the CMAC sub-environment, the red QMIX agent performs well in scenario 2, better in scenario 1, but performs poorly in scenario 0, which is a counter-intuitive result.

In the AMAC environment, due to the asymmetric property, the win rate of the red QMIX agent is stable at 1, and the blue QMIX agent has an improved win rate in scenario 1 after a period of training but performs poorly in scenario 0 and scenario 2.

In the SRAMC sub-environment, the QMIX agent performs well as both the red and blue sides in scenario 1, poor in scenario 0, and very poor in scenario 2.

\section{Conclusion}
In this paper, we have introduced the WarGame Challenge (WGC), a partially observable multi-agent cooperation benchmark environment for more like real world game challenge, which includes five sub-environments and with a easy-to-use and flexible framework. In the future, we plan to expand WGC with more challenging scenarios, such as increasing communication between operators for MARL communication tasks, adding more rules, offering additional setting files, and improving the interface to make it a more universal environment. We welcome contributions from the community to WGC and hope that it will become a standard benchmark for measuring progress in MARL.

\bibliographystyle{ieeetr}
\bibliography{ref}

\begin{thebibliography}{10}

\bibitem{candela2022transferring}
E.~Candela, L.~Parada, L.~Marques, T.-A. Georgescu, Y.~Demiris, and
  P.~Angeloudis, ``Transferring multi-agent reinforcement learning policies for
  autonomous driving using sim-to-real,'' in {\em 2022 IEEE/RSJ International
  Conference on Intelligent Robots and Systems (IROS)}, pp.~8814--8820, IEEE,
  2022.

\bibitem{li2022v2x}
Y.~Li, D.~Ma, Z.~An, Z.~Wang, Y.~Zhong, S.~Chen, and C.~Feng, ``V2x-sim:
  Multi-agent collaborative perception dataset and benchmark for autonomous
  driving,'' {\em IEEE Robotics and Automation Letters}, vol.~7, no.~4,
  pp.~10914--10921, 2022.

\bibitem{zhou2022multi}
W.~Zhou, D.~Chen, J.~Yan, Z.~Li, H.~Yin, and W.~Ge, ``Multi-agent reinforcement
  learning for cooperative lane changing of connected and autonomous vehicles
  in mixed traffic,'' {\em Autonomous Intelligent Systems}, vol.~2, no.~1,
  p.~5, 2022.

\bibitem{toghi2022social}
B.~Toghi, R.~Valiente, D.~Sadigh, R.~Pedarsani, and Y.~P. Fallah, ``Social
  coordination and altruism in autonomous driving,'' {\em IEEE Transactions on
  Intelligent Transportation Systems}, vol.~23, no.~12, pp.~24791--24804, 2022.

\bibitem{bettini2023heterogeneous}
M.~Bettini, A.~Shankar, and A.~Prorok, ``Heterogeneous multi-robot
  reinforcement learning,'' {\em arXiv preprint arXiv:2301.07137}, 2023.

\bibitem{johnson2022multi}
D.~Johnson, G.~Chen, and Y.~Lu, ``Multi-agent reinforcement learning for
  real-time dynamic production scheduling in a robot assembly cell,'' {\em IEEE
  Robotics and Automation Letters}, vol.~7, no.~3, pp.~7684--7691, 2022.

\bibitem{liang2022multi}
Z.~Liang, J.~Cao, S.~Jiang, D.~Saxena, J.~Chen, and H.~Xu, ``From multi-agent
  to multi-robot: A scalable training and evaluation platform for multi-robot
  reinforcement learning,'' {\em arXiv preprint arXiv:2206.09590}, 2022.

\bibitem{bettini2022vmas}
M.~Bettini, R.~Kortvelesy, J.~Blumenkamp, and A.~Prorok, ``Vmas: A vectorized
  multi-agent simulator for collective robot learning,'' {\em arXiv preprint
  arXiv:2207.03530}, 2022.

\bibitem{qiu2023multi}
C.~Qiu, Z.~Wu, J.~Wang, M.~Tan, and J.~Yu, ``Multi-agent reinforcement learning
  based stable path tracking control for a bionic robotic fish with reaction
  wheel,'' {\em IEEE Transactions on Industrial Electronics}, 2023.

\bibitem{berner2019dota}
C.~Berner, G.~Brockman, B.~Chan, V.~Cheung, P.~Debiak, C.~Dennison, D.~Farhi,
  Q.~Fischer, S.~Hashme, C.~Hesse, {\em et~al.}, ``Dota 2 with large scale deep
  reinforcement learning,'' {\em arXiv preprint arXiv:1912.06680}, 2019.

\bibitem{jaderberg2019human}
M.~Jaderberg, W.~M. Czarnecki, I.~Dunning, L.~Marris, G.~Lever, A.~G.
  Castaneda, C.~Beattie, N.~C. Rabinowitz, A.~S. Morcos, A.~Ruderman, {\em
  et~al.}, ``Human-level performance in 3d multiplayer games with
  population-based reinforcement learning,'' {\em Science}, vol.~364, no.~6443,
  pp.~859--865, 2019.

\bibitem{vinyals2019alphastar}
O.~Vinyals, I.~Babuschkin, J.~Chung, M.~Mathieu, M.~Jaderberg, W.~M. Czarnecki,
  A.~Dudzik, A.~Huang, P.~Georgiev, R.~Powell, {\em et~al.}, ``Alphastar:
  Mastering the real-time strategy game starcraft ii,'' {\em DeepMind blog},
  vol.~2, p.~20, 2019.

\bibitem{smac}
M.~Samvelyan, T.~Rashid, C.~S. De~Witt, G.~Farquhar, N.~Nardelli, T.~G. Rudner,
  C.-M. Hung, P.~H. Torr, J.~Foerster, and S.~Whiteson, ``The starcraft
  multi-agent challenge,'' {\em arXiv preprint arXiv:1902.04043}, 2019.

\bibitem{gronauer2022multi}
S.~Gronauer and K.~Diepold, ``Multi-agent deep reinforcement learning: a
  survey,'' {\em Artificial Intelligence Review}, vol.~55, no.~2, pp.~895--943,
  2022.

\bibitem{lowe}
R.~Lowe, Y.~Wu, A.~Tamar, J.~Harb, P.~Abbeel, and I.~Mordatch, ``Multi-agent
  actor-critic for mixed cooperative-competitive environments,'' {\em arXiv
  preprint arXiv:1706.02275}, 2017.

\bibitem{resnick2018pommerman}
C.~Resnick, W.~Eldridge, D.~Ha, D.~Britz, J.~Foerster, J.~Togelius, K.~Cho, and
  J.~Bruna, ``Pommerman: A multi-agent playground,'' {\em arXiv preprint
  arXiv:1809.07124}, 2018.

\bibitem{kurach2019google}
K.~Kurach, A.~Raichuk, P.~Stanczyk, M.~Zajac, O.~Bachem, L.~Espeholt,
  C.~Riquelme, D.~Vincent, M.~Michalski, O.~Bousquet, {\em et~al.}, ``Google
  research football: A novel reinforcement learning environment,'' {\em arXiv
  preprint arXiv:1907.11180}, 2019.

\bibitem{jia2020fever}
H.~Jia, Y.~Hu, Y.~Chen, C.~Ren, T.~Lv, C.~Fan, and C.~Zhang, ``Fever
  basketball: A complex, flexible, and asynchronized sports game environment
  for multi-agent reinforcement learning,'' {\em arXiv preprint
  arXiv:2012.03204}, 2020.

\bibitem{baker2019emergent}
B.~Baker, I.~Kanitscheider, T.~Markov, Y.~Wu, G.~Powell, B.~McGrew, and
  I.~Mordatch, ``Emergent tool use from multi-agent interaction,'' {\em Machine
  Learning, Cornell University}, 2019.

\bibitem{rashid2018qmix}
T.~Rashid, M.~Samvelyan, C.~Schroeder, G.~Farquhar, J.~Foerster, and
  S.~Whiteson, ``Qmix: Monotonic value function factorisation for deep
  multi-agent reinforcement learning,'' in {\em International Conference on
  Machine Learning}, pp.~4295--4304, PMLR, 2018.

\bibitem{son2019qtran}
K.~Son, D.~Kim, W.~J. Kang, D.~E. Hostallero, and Y.~Yi, ``Qtran: Learning to
  factorize with transformation for cooperative multi-agent reinforcement
  learning,'' in {\em International Conference on Machine Learning},
  pp.~5887--5896, PMLR, 2019.

\bibitem{foerster2018counterfactual}
J.~Foerster, G.~Farquhar, T.~Afouras, N.~Nardelli, and S.~Whiteson,
  ``Counterfactual multi-agent policy gradients,'' in {\em Proceedings of the
  AAAI Conference on Artificial Intelligence}, vol.~32, 2018.

\bibitem{yang2018mean}
Y.~Yang, R.~Luo, M.~Li, M.~Zhou, W.~Zhang, and J.~Wang, ``Mean field
  multi-agent reinforcement learning,'' in {\em International Conference on
  Machine Learning}, pp.~5571--5580, PMLR, 2018.

\bibitem{leibo}
J.~Z. Leibo, V.~Zambaldi, M.~Lanctot, J.~Marecki, and T.~Graepel, ``Multi-agent
  reinforcement learning in sequential social dilemmas,'' {\em arXiv preprint
  arXiv:1702.03037}, 2017.

\bibitem{peng2021facmac}
B.~Peng, T.~Rashid, C.~Schroeder~de Witt, P.-A. Kamienny, P.~Torr,
  W.~B{\"o}hmer, and S.~Whiteson, ``Facmac: Factored multi-agent centralised
  policy gradients,'' {\em Advances in Neural Information Processing Systems},
  vol.~34, pp.~12208--12221, 2021.

\bibitem{zhang2019cityflow}
H.~Zhang, S.~Feng, C.~Liu, Y.~Ding, Y.~Zhu, Z.~Zhou, W.~Zhang, Y.~Yu, H.~Jin,
  and Z.~Li, ``Cityflow: A multi-agent reinforcement learning environment for
  large scale city traffic scenario,'' in {\em The world wide web conference},
  pp.~3620--3624, 2019.

\bibitem{zha2021douzero}
D.~Zha, J.~Xie, W.~Ma, S.~Zhang, X.~Lian, X.~Hu, and J.~Liu, ``Douzero:
  Mastering doudizhu with self-play deep reinforcement learning,'' in {\em
  International Conference on Machine Learning}, pp.~12333--12344, PMLR, 2021.

\bibitem{moy2019application}
G.~Moy and S.~Shekh, ``The application of alphazero to wargaming,'' in {\em
  Australasian Joint Conference on Artificial Intelligence}, pp.~3--14,
  Springer, 2019.

\bibitem{oliehoek2016concise}
F.~A. Oliehoek and C.~Amato, {\em A concise introduction to decentralized
  POMDPs}.
\newblock Springer, 2016.

\bibitem{halpern2007computer}
J.~Y. Halpern, ``Computer science and game theory: A brief survey,'' {\em arXiv
  preprint cs/0703148}, 2007.

\end{thebibliography}

\end{document}